\documentclass{ieeeaccess}
\usepackage[numbers, square, comma, sort&compress]{natbib}
\usepackage{amsmath,amssymb,amsfonts}
\usepackage[lined, linesnumbered, ruled]{algorithm2e}
\usepackage{graphicx}
\usepackage{subfigure}
\usepackage{textcomp}
\usepackage{caption}
\usepackage{multicol, multirow}
\usepackage{color,soul}
\usepackage{xcolor,colortbl}
\usepackage[none]{hyphenat}
\def\BibTeX{{\rm B\kern-.05em{\sc i\kern-.025em b}\kern-.08emT\kern-.1667em\lower.7ex\hbox{E}\kern-.125emX}}

\begin{document}

\history{Date of publication xxxx 00, 0000, date of current version xxxx 00, 0000.} \doi{0000/ACCESS.0000.DOI}

\title{HEVC Watermarking Techniques for Authentication and Copyright Applications: Challenges and Opportunities}

\author{
	\uppercase{Ali A. Elrowayati}\authorrefmark{1*} 	\IEEEmembership{Member, IEEE},
	\uppercase{MOHAMED A. ALRSHAH}	\authorrefmark{2*} 	\IEEEmembership{Senior Member, IEEE},
	\uppercase{M.F.L. Abdullah} \authorrefmark{3*} \IEEEmembership{Senior Member, IEEE}, and 
	\uppercase{ROHAYA LATIP} \authorrefmark{2} \IEEEmembership{Member, IEEE}
}

\address[1]{Department of Electronic Engineering, the College of Industrial Technology, Misurata, Libya}
\address[2]{Department of Communication Technology \& Network, Faculty of Computer Science \& IT, Universiti Putra Malaysia, 43400, Serdang, Selangor, Malaysia}
\address[3]{Faculty of Electrical and Electronic Engineering, Universiti Tun Hussein Onn Malaysia, 86400, Parit Raja, Johor, Malaysia}

\tfootnote{This work has been financially supported by Universiti Putra Malaysia (UPM). Moreover, the technical and logistic parts of this work have been achieved by a research team formed of members from the College of Industrial Technology - Libya, Universiti Putra Malaysia, and Universiti Tun Hussein Onn Malaysia.}

\markboth
{Ali A. Elrowayati \headeretal: HEVC Watermarking Techniques for Authentication and Copyright Applications: Challenges and Opportunities}
{Ali A. Elrowayati \headeretal: HEVC Watermarking Techniques for Authentication and Copyright Applications: Challenges and Opportunities}

\corresp{Corresponding authors: Ali A. Elrowayati (elrowayati@yahoo.com) and Mohamed A. Alrshah (mohamed.asnd@gmail.com), M.F.L. Abdullah (faiz@uthm.edu.my)}
	
\begin{abstract}
Recently, High-Efficiency Video Coding (HEVC/H.265) has been chosen to replace previous video coding standards, such as H.263 and H.264. Despite the efficiency of HEVC, it still lacks reliable and practical functionalities to support authentication and copyright applications. In order to provide this support, several watermarking techniques have been proposed by many researchers during the last few years. However, those techniques are still suffering from many issues that need to be considered for future designs. In this paper, a Systematic Literature Review (SLR) is introduced to identify HEVC challenges and potential research directions for interested researchers and developers. The time scope of this SLR covers all research articles published during the last six years starting from January 2014 up to the end of April 2020. Forty-two articles have met the criteria of selection out of 343 articles published in this area during the mentioned time scope. A new classification has been drawn followed by an identification of the challenges of implementing HEVC watermarking techniques based on the analysis and discussion of those chosen articles. Eventually, recommendations for HEVC watermarking techniques have been listed to help researchers to improve the existing techniques or to design new efficient ones.
\end{abstract}

\begin{keywords}
	Authentication, Copyright, HEVC, H.265, SLR, Systematic Review, Video, Watermarking 
\end{keywords}

\titlepgskip=-15pt

\maketitle

\section{Introduction}
In the last decade, video coding standards have been advancing to provide better data compression while maintaining high-quality visual resolution. Further, authentication and copyright protection have become a key interest to protect video content. However, accessing and tampering video contents became an easy task that disturbs the process of authentication and copyright protection due to the availability of video content and advanced video editing tools on the Internet. This increases the necessity of finding solutions that can protect copyrights, detect, and localize video tampering \cite{1}. Thus, integrated solutions in video compression standards to provide copyright protection, and authentication are significantly needed, in which these solutions must have the ability to validate authentication, copyright information, and content integrity.

The previous video codecs, such as H.263 and H.264, have been successfully protected by many watermarking techniques \cite{2, 3, 4, 5}. However, the High-Efficiency Video Coding (HEVC) standard is poising to replace those watermarking techniques, despite their success, in order to maintain the flexibility, reliability, and robustness of the HEVC \cite{6}. Furthermore, applying watermarking techniques for HEVC video protection is relatively a new research trend that is full of legal challenges such as copyright protection, ownership proofing, tampering detection, and authentication \cite{1, 7, 8}. 

Although HEVC watermarking techniques have been existing for a few years, this field suffers from the absence of Systematic Literature Review (SLR). As well-known, SLR papers are very necessary to ease obtaining the latest updates, such as the open issues and research gaps, in a specific topic to save the time and effort of the researchers who are willing to contribute to the field. Indeed, the main contributions of this paper are:
\begin{itemize}
	\item Highlighting the necessity of using watermarking with the HEVC.
	\item Presenting a new classification of the existing and/or possible watermarking techniques. 
	\item Disclosing common challenges and issues of integrating watermarking techniques into the HEVC codec.
	\item Summarizing the potential applications of authentication and copyrighting that need to be considered in future HEVC watermarking techniques. 
	\item Answering several common critical questions related to HEVC watermarking in order to open the door for new trends and domains in this area.
\end{itemize}

The remainder of this paper has been structured as follows: Section \ref{meth} presents the method and procedure of this SLR; including the study questions, searching strategy and selection criteria. Section \ref{analysis} shows statistics, classifications, analysis, and discussion with highlighting the challenges and open issues in this area. Finally, Section \ref{conc} concludes the work, and Section \ref{futur} presents the future directions, respectively.

\section{SLR Method and Procedure} \label{meth}
This section describes the followed procedure in this paper to present an unbiased coverage of studied literature. The process includes: defining the directive study questions, determining the search strategy, and assessing articles based on the selection criteria.

\subsection{Defining the Directive Study Questions}
Based on gaps found in the recent articles related to the scope of this research, the following eight important questions (Qs) are going to be answered in this paper: 
\begin{itemize}
	\item[\textbf{Q1.}] What are the main differences among data hiding, steganography, cryptography, and watermarking terms?
	\item[\textbf{Q2.}] Is there still a need for HEVC watermarking techniques to support copyright and authentication applications?
	\item[\textbf{Q3.}] What are the current watermarking techniques that are used or can be applied for the HEVC codec to provide authentication and copyright functionalities?
	\item[\textbf{Q4.}] What are the possible options to implement watermarking techniques into the HEVC codec for authentication and copyright applications?
	\item[\textbf{Q5.}] What are the main watermark zone selection criteria that can be applied to design efficient HEVC watermarking techniques for authentication and copyright applications?
	\item[\textbf{Q6.}] What are the common metrics used to evaluate the performance of video watermarking techniques?
	\item[\textbf{Q7.}] What are the main challenges in HEVC video watermarking?
	\item[\textbf{Q8.}] Are there any real-time HEVC video watermarking techniques implemented on hardware platform?
\end{itemize}

\textbf{Q1} aims to illustrate the main difference among data hiding, steganography, cryptography, and watermarking terms. \textbf{Q2} aims at highlighting the necessity of watermarking techniques for authenticating and copyright-protecting HEVC videos. \textbf{Q3} investigates the existing watermarking techniques and their applicability to the HEVC standard. \textbf{Q4} discusses the possible options of the HEVC video watermarking techniques for authentication and copyright applications. \textbf{Q5}  discusses the main watermark zone selection criteria that can be applied for designing HEVC video watermarking techniques for authentication and copyright applications. In \textbf{Q6}, the common metrics used to evaluate the performance of general video watermarking are discussed. \textbf{Q7} presents the most critical challenges of the existing HEVC watermarking techniques, including common and security challenges for both authentication and copyright applications. Finally, \textbf{Q8} finds whether real-time HEVC video watermarking techniques are implemented on hardware platforms. 

\subsection{Searching Strategy and Selection Criteria}
In this research, three criteria have been used for paper selection: 
\begin{itemize}
	\item The scope of this review is limited to published research on HEVC video watermarking techniques for authentication and copyright applications. 
	\item The time scope of this SLR covers all research papers published in period from January 2014 up to the end of April 2020. 
	\item General keywords and their alternative spellings and synonyms have been used to search in the digital libraries. These keywords have been used with the Boolean (AND) and (OR) which are used to connect among the keywords and their alternative spellings and synonyms as shown in Table \ref{keywords}.
\end{itemize}

The searching process targets the well-known \textit{ScienceDirect, Scopus, IEEExplore, Web of Science} and the other academic digital libraries that contain peer-reviewed journal articles, conference proceedings, and book chapters. Multiple academic tools, such as Google Scholar engine and EndNote X7.5, have been used for gathering and obtaining a comprehensive list of relevant articles. These tools have been used to perform an automatic search in the identified resources using the most appropriate search strings, keywords, and synonyms, as in Table \ref{keywords}. Initially, this search strategy produces lists of related and interesting articles including many duplicated and redundant items. For this reason, avoiding duplication and redundancy during the articles selection process was a significant step to consolidate similar articles to obtain a list of the most relevant and unique articles. 

\begin{table}[!h]
	\caption{Searching Keywords, Synonyms, and Boolean Operators} \label{keywords}
	\begin{center}{\scriptsize 
			\begin{tabular}{|c|} 	\hline \\		
				\begin{tabular}{|c|c|c|c|}		\hline
					High Efficiency Video Coding & OR & \multicolumn{2}{c|}{\begin{tabular}{c|c|c} HEVC & OR & H.265 \end{tabular}} \\  \hline
					\multicolumn{4}{c}{}												 \\ 
					\multicolumn{4}{c}{AND}                                  	 		 \\
					\multicolumn{4}{c}{}												 \\ \hline
					Watermarking                    & OR & Robust Watermarking  	& OR \\ \hline
					Fragile Watermarking            & OR & Readable Watermarking    & OR \\ \hline
					Detectable Watermarking         & OR & Zero-Watermarking        & OR \\ \hline
					Information Hiding              & OR & Data Embedding           & OR \\ \hline
					Security                        & OR & Authentication           & OR \\ \hline
					Copyright                       & OR & Video Forensics          & OR \\ \hline
					Double Compression Detection    & OR & Re-compression Detection & OR \\ \hline
					Tampering Detection				& OR & Semi-fragile				& \cellcolor[HTML]{333333}\\ \hline
					
				\end{tabular}\\ \\ \hline
			\end{tabular}
		}
	\end{center}
\end{table}

Specifically, all obtained articles are filtered using standard search procedures and guidelines as in \cite{19}. More specifically, the inclusion and exclusion criteria, shown in Table \ref{tbl:tbl1}, are applied for each article to choose the most relevant ones.
\begin{table}[!h]
	\caption{Inclusion and exclusion criteria}\label{tbl:tbl1}
	\begin{center}
	\begin{tabular}{m{3.8cm}m{3.8cm}}\hline
		\textbf{Inclusion criteria}                           										& \textbf{Exclusion criteria}   \\\hline
		\rowcolor[HTML]{EFEFEF} Directly related to the main topic      							& Irrelevant to the main topic  \\
		Presents HEVC watermarking technique, method and results 									& Published in a preliminary conference\\
		\rowcolor[HTML]{EFEFEF}Presents experimental dataset, evaluation metrics and discussion		& Review papers or unpublished papers \\
		Answers the presented research questions 													& Incomplete or includes hidden parts\\
		\rowcolor[HTML]{EFEFEF} Written in English                      							& Not written in English			\\\hline           
	\end{tabular}
	\end{center}
\end{table}

After filtering the articles using the inclusion and exclusion criteria, the obtained list of articles is considered the final comprehensive Primary Study List (PSL), which includes the most relevant and related articles without overlapping, redundancy and/or duplication.

\section{Analysis and Discussion} \label{analysis}
As a result of the aforementioned searching strategy and selection criteria, the obtained PSL has only included 42 articles selected based on the inclusion and exclusion criteria from a total of 343 articles, as shown in Table \ref{tbl:tbl2}. Specifically, there were 298 articles excluded from the list since they were not satisfying the inclusion criteria. In this section, the metadata of the obtained PSL articles has been analyzed to present some useful statistics. Then, the questions of the study have been answered based on the analysis and criticism of the PSL articles.

\begin{figure} [!h]
	\centering
	\begin{center}
		\subfigure [Number of publications per year based on publication type]
		{
			\includegraphics[width=0.95\linewidth]{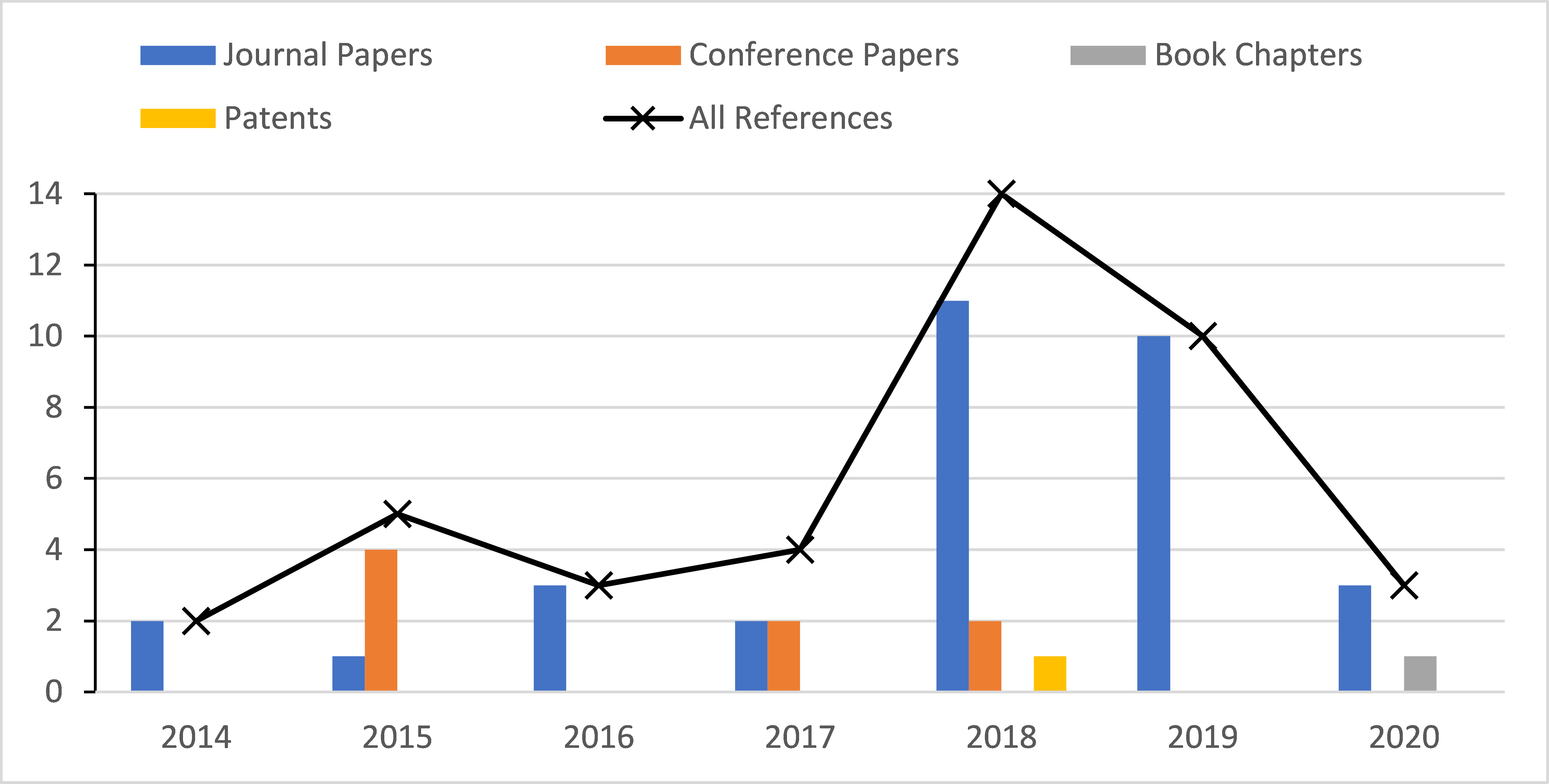}
			\label{fig:fig1a}
		}
		\subfigure [Percentage of publications based on publication type]
		{
			\includegraphics[width=0.95\linewidth]{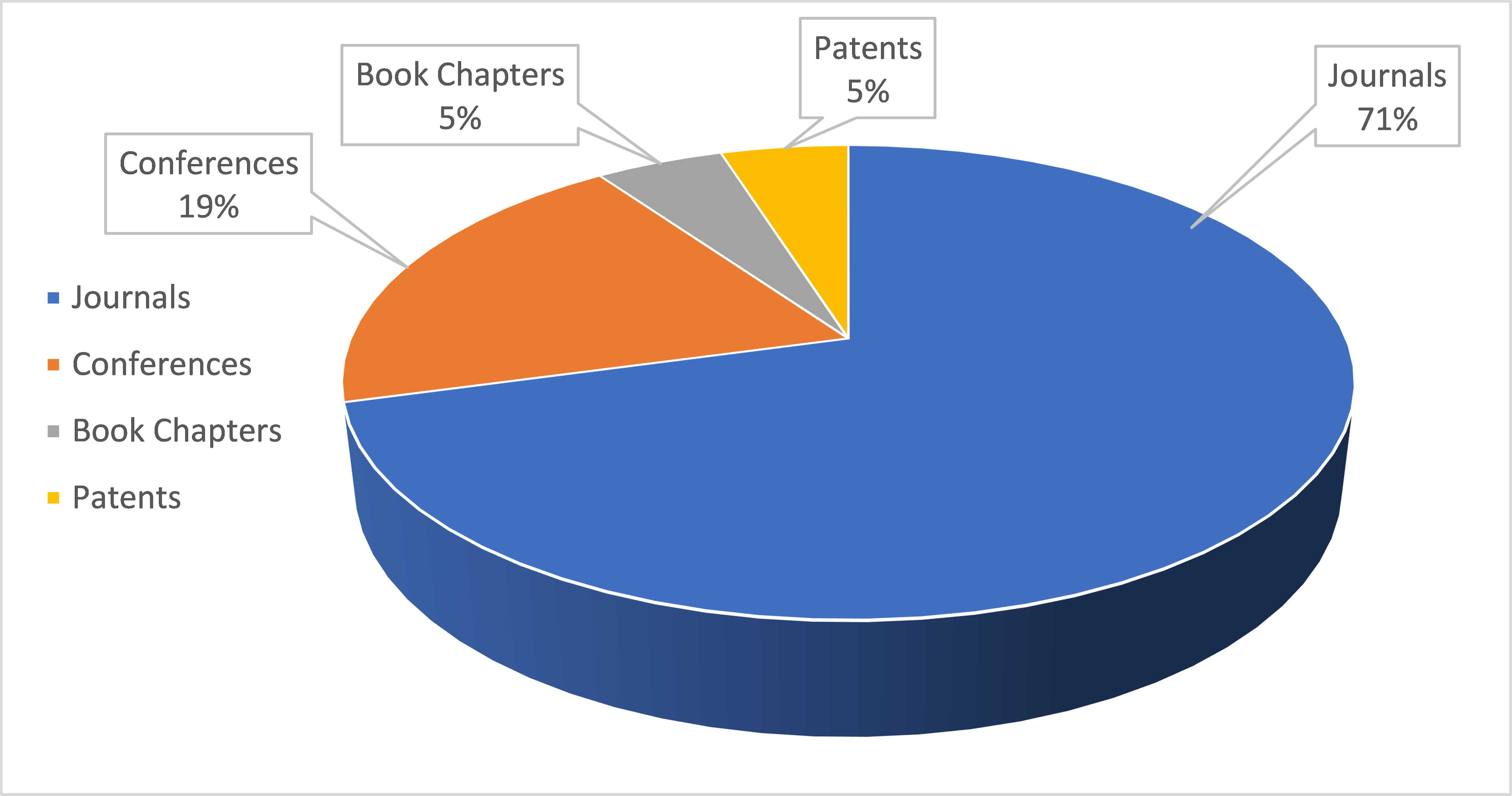}
			\label{fig:fig1b}
		}
		\subfigure [Percentage of publications based on country]
		{
			\includegraphics[width=0.95\linewidth]{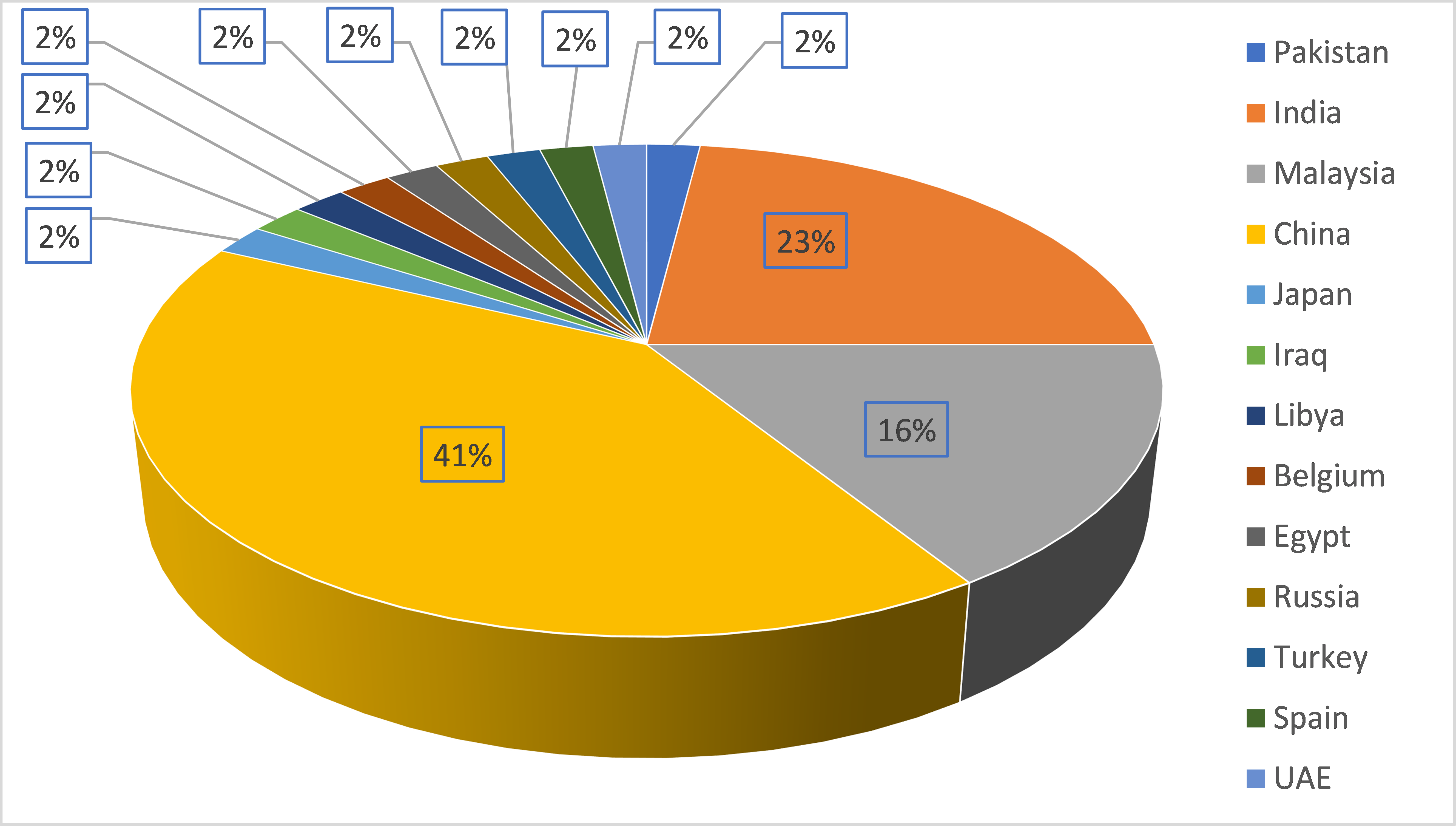}
			\label{fig:fig1c}
		}
	\end{center}
	\caption{PSL statistics based on type, year, and country.}
	\label{fig:fig1}
\end{figure}

Fig. \ref{fig:fig1a}, shows the distribution of the PSL articles over the last five years based on the article's type. In general, it is clear that the number of publications in this field is very limited. However, it is also clear that the number of publications has been dramatically increased in the last two years. The first data embedding technique, which inserts and hides data into HEVC video content to protect copyright information, has been published in 2014 as a journal paper (reference \cite{16}). As for the first fragile HEVC watermarking technique to support authentication features, it has been published also in 2014 as a journal paper  (reference \cite{35}). As for the first robust HEVC watermarking technique against the re-compression attack, it has been published under a conference proceeding in 2015 (reference \cite{36}). To conclude, it is clearly noticed that HEVC watermarking received increasing attention in the year 2017 onwards.

As for Fig. \ref{fig:fig1b}, it shows the percentage of all article types in the PSL; 71\% journal papers, 19\% conference papers, 5\% chapter of books, and 5\% patents. Therefore, the researchers in this field need to focus more on publishing conference papers to enrich the discussion and ideas about the HEVC. As well as, they need to publish their full research papers in high-impact journals.

Regarding Fig. \ref{fig:fig1c}, it shows the distribution of the PSL articles based on the author's country. It is very clear that China, India, and Malaysia have been the most contributing countries in terms of published research articles in this area with 41\%, 23\%, and 16\%, respectively. As for the remaining 20\%, it has been equally distributed among the other countries; Belgium, Libya, Iraq, Pakistan, Japan, Egypt, Spain, Russia, Turkey, and UAE; with 2\% for each country.

\begin{table*}[!h]
	\caption{The obtained PSL articles selected based on the inclusion and exclusion criteria} \label{tbl:tbl2}
	\begin{center}\scriptsize
		\begin{tabular}{m{0.4cm}m{2.9cm}m{9.85cm}m{1.65cm}m{0.75cm}} \hline
			No. & Author(s), Year, Ref\#              & Title            & Reference Type         & Country  \\\hline
			\rowcolor[HTML]{EFEFEF}1   & Chang et al., 2014 \cite{16}      & A DCT/DST-based error propagation-free data hiding algorithm for HEVC intra-coded frames.                                    & Journal Article        & China         \\
			2   & Swati et al., 2014 \cite{35}     & Watermarking scheme for high efficiency video coding (HEVC).                                                                 & Journal Article        & Pakistan        \\
			\rowcolor[HTML]{EFEFEF}3   & Gaj et al., 2015 \cite{36}        & A robust watermarking scheme against re-compression attack for H.265/HEVC.                                                  & Conference Proceedings & India           \\
			4   & Abdullah et al., 2015 \cite{10}   & Recent methods and techniques in video watermarking and their applicability to the next generation video codec.              & Journal Article        & Libya \\
			\rowcolor[HTML]{EFEFEF}5   & Huang et al., 2015 \cite{26}     & Detection of double compression for HEVC videos based on the co-occurrence matrix of DCT coefficients.                       & Conference Proceedings & China           \\
			6   & Ogawa, \& Ohtake, 2015 \cite{37}                            & Watermarking for HEVC/H.265 stream.                                                                                         & Conference Proceedings & Japan           \\
			\rowcolor[HTML]{EFEFEF}7   & Tew et al., 2015 \cite{11}              & HEVC video authentication using data embedding technique.                                                                    & Conference Proceedings & Malaysia        \\
			8   & Dutta \& Gupta, 2016 \cite{12}   & A robust watermarking framework for high efficiency video coding (HEVC) – Encoded video with blind extraction process.       & Journal Article        & India           \\
			\rowcolor[HTML]{EFEFEF}9   & Elrowayati et al., 2016 \cite{13}    & Robust HEVC video watermarking scheme based on repetition-BCH syndrome code.                & Journal Article        & Malaysia        \\
			10  & Tew et al., 2016 \cite{14}       & Multi-layer authentication scheme for HEVC video based on embedded statistics.                                               & Journal Article        & Malaysia        \\
			\rowcolor[HTML]{EFEFEF}11  & Dutta \& Gupta, 2017 \cite{8}   & An efficient framework for compressed domain watermarking in P frames of high-efficiency video coding (HEVC)--Encoded video. & Journal Article        & India           \\
			12  & Elrowayati et al., 2017 \cite{7}       & Tampering detection of double-compression with the same quantization parameter in HEVC video streams.                        & Conference Proceedings & Malaysia        \\
			\rowcolor[HTML]{EFEFEF}13  & Gaj et al., 2017 \cite{38}      & Drift-compensated robust watermarking algorithm for H. 265/HEVC video stream.             & Journal Article         & India           \\
			14  &  Xu et al., 2017 \cite{27}      & Zero-watermarking registration and detection method for HEVC video streaming against requantization transcoding.             & Google Patents         & China           \\
			\rowcolor[HTML]{EFEFEF}15  & Jiang et al., 2018 \cite{95}      & HEVC Double Compression Detection Based on SN-PUPM Feature.            				&Conference Proceedings         & China           \\
			16  & Jo et al., 2018 \cite{15}         & A Reversible Watermarking Algorithm in the Lossless Mode of HEVC.                                                            & Journal Article        & China           \\
			\rowcolor[HTML]{EFEFEF}17  & Joa et al., 2018 \cite{33}        & A watermarking method by modifying QTCs for HEVC.                                                                            & Journal Article        & China           \\
			18  & Kaur et al., 2018 \cite{34}       & An efficient watermarking scheme for enhanced high efficiency video coding/H.265.                                           & Journal Article        & India           \\
			\rowcolor[HTML]{EFEFEF}19  & Liu et al., 2018 \cite{17}        & A robust and improved visual quality data hiding method for HEVC.                                                            & Journal Article        & China           \\
			20  & Mohammed \& Ali, 2018 \cite{18}           & Robust video watermarking scheme using high efficiency video coding attack.                                                  & Journal Article        & Iraq            \\
			\rowcolor[HTML]{EFEFEF}21  & Tew et al., 2018 \cite{1}        & Separable authentication in encrypted HEVC video.                                                                            & Journal Article        & Malaysia        \\
			22	& Mareen et al., 2018 \cite{31}	  & A Novel Video Watermarking Approach Based on Implicit Distortions															 & Journal Article 		  &
			Belgium			\\
			\rowcolor[HTML]{EFEFEF}23	& Liu et al., 2018 \cite{32}		  & Hiding Bitcoin Transaction Information Based on HEVC																		 & Conference Proceedings &
			China            \\
			24 & Joshi et al., 2018a \cite{47}		  & Real-Time Implementation of Blind and Robust Watermarking for HEVC Video Coding.																		 & Conference Proceedings &
			India	         \\
			\rowcolor[HTML]{EFEFEF}25	& Joshi, 2018b \cite{43}		  & VLSI Implementation of Video Watermarking for Secure HEVC Coding Standard.																		 & Journal Article &
			India            \\
			26	& Li, Wang \& Xu, 2018	\cite{46}	  & Detection of double compression in HEVC videos based on TU size and quantized DCT coefficients.																		 & Journal Article &
			China 	        \\
			\rowcolor[HTML]{EFEFEF}27	& Liang et al., 2018 \cite{45}		  & Detection of double compression for HEVC videos with a fake bitrate.																		 & Journal Article &
			China          \\
			28	& Wang et al., 2018 \cite{101}		  & Anti-HEVC Recompression Video Watermarking Algorithm Based on the All Phase Biorthogonal Transform and SVD.																		 & Journal Article &
			China          \\			
			\rowcolor[HTML]{EFEFEF}29	& El-Shafai et al., 2019 \cite{48}		  & Security of 3D-HEVC transmission based on fusion and watermarking techniques.																		 & Journal Article &
			Egypt 	        \\
			30	& X.Yu, et al., 2019 \cite{49}		  & A Hybrid Transforms-Based Robust Video Zero-Watermarking Algorithm for Resisting High Efficiency Video Coding Compression.																		 & Journal Article &
			China          \\
			\rowcolor[HTML]{EFEFEF}31 & L.Yu, et al., 2019 \cite{50}		  & HEVC double compression detection under different bitrates based on TU partition type.																		 & Journal Article &
			China 	        \\
			32	& Shanableh, 2019 \cite{51}		  & Data Embedding in HEVC Video by Modifying the Partitioning of Coding Units.																		 & Journal Article &
			UAE          \\
			\rowcolor[HTML]{EFEFEF}33  & Kaur et al., 2019a \cite{52}		  & An efficient watermarking scheme for enhanced high efficiency video coding/H.265.																		 & Journal Article &
			India 	        \\
			34	& Jiang, et al., 2019a \cite{53}		  & Detection of Double Compressed HEVC Videos Using GOP-Based PU Type Statistics.																		 & Journal Article &
			China           \\
			\rowcolor[HTML]{EFEFEF}35	& Kaur et al., 2019b \cite{83}	  & An efficient authentication scheme for high efficiency video coding/H.265.																		 & Journal Article &
			India 	        \\
			36	& Fang et al., 2019	\cite{84}	  & Detection of HEVC Double Compression with Different Quantization Parameters Based on Property of DCT Coefficients and TUs.																		 & Journal Article &
			China			\\
			\rowcolor[HTML]{EFEFEF}37	& Jang et al., 2019 \cite{90}	  & Biological Viral Infection Watermarking Architecture of MPEG/H. 264/AVC/HEVC																		 & Journal Article &
			China	        \\
			38	& Jiang, et al., 2019 b	\cite{89}	  & Detection of HEVC Double Compression with the Same Coding Parameters Based on Analysis of Intra-Coding Quality Degradation Process		& Journal Article &
			China			\\
			\rowcolor[HTML]{EFEFEF}39	& Galiano  et al., 2020 \cite{94}	  & Efficient embedding and retrieval of information for high-resolution videos coded with HEVC																		 & Journal Article &
			Spain	        \\
			40	& Favorskaya, et al., 2019 \cite{93}	  & Authentication and Copyright Protection of Videos Under Transmitting Specifications		& Chapter of Book &
			Russia			\\
			\rowcolor[HTML]{EFEFEF}41	& Konyar et al., 2020 \cite{92}	  & Matrix encoding-based high-capacity and high-fidelity reversible data hiding in HEVC																		 & Journal Article &
			Turkey	        \\
			42	& Gaj et al., 2020 \cite{97}	  & Prediction mode based H. 265/HEVC video watermarking resisting re-compression attack																		 & Journal Article &
			India	        \\	\hline				
		\end{tabular}
	\end{center}
\end{table*}

\subsection{Q1: What are the main differences among data hiding, steganography, cryptography, and watermarking terms?}
Data hiding is a general term comprising a wide range of content problems beyond embedding messages, such as steganography, cryptography, and watermarking. The differences among these terms are fundamental and based on different requirements, designs, and technical solutions. The steganography is a word derived from the Greek word (Steganographia), where \textit{steganos} means “\textit{covered}” and \textit{graphia} means “\textit{writing}”. Steganography is an age-old technique of hiding plain data within another hosting file as a concealed communication, which allows exchanging information without arousing any suspicion. As for cryptography, it is used to protect the hidden message by encrypting its content. More specifically, the third parties in the steganography should not know about the existence of the plain text in the hosting media, while in the encryption, the existence of the protected data is known for the public. As for the digital watermarking, it is the process of embedding information into digital media, no matter it is perceptible or imperceptible to the third parties. For instance, most broadcasting providers use perceptible watermarks that are visible for the public to protect the copyright of the broadcasted content. Furthermore, the watermarks can be also hidden or encrypted into the hosting video for authentication purposes such as preventing unauthorized use for the broadcasted content \cite{20, 64, 74, 69}

There are four fundamental differences between steganography and watermarking:
\begin{enumerate}
	\item Watermarks have to be resistant against possible attacks unlike messages in steganography. 
	\item Watermarks could be visible or invisible, while messages in steganography must be hidden.
	\item In steganography, there is no relationship between the host file and the message, while the watermark uses a relevant message to protect or maintain the ownership and/or the authentication of the host file \cite{81}.
	\item Based on the number of senders and receivers, steganography is usually a one-to-one application, while watermarking is usually a one-to-many application \cite{76}.
\end{enumerate}

\subsection{Q2: Is there still a need for HEVC watermarking techniques to support copyright and authentication applications?}
Fig. \ref{fig:fig1a} shows that HEVC watermarking is an open area for research due to the clear increase in the number of publications in the last few years. In addition, the HEVC video standard and its watermarking for authentication and copyright purposes become very widely used by video streaming industries, such as Netflix, HBO Go, and others. 

The only way to do video watermarking is by embedding/extracting the watermark into/from video content during the encoding and decoding processes, respectively, as the video streaming companies exactly do. Moreover, many encoding tools, video editing software, and some video capturing cards now provide functionalities of embedding/extracting watermarks into/from HEVC encoded video streams. Therefore, the integrity check, tampering detection, high visual quality assurance, bitrate control, and readability of watermarks on the receiver side, are significantly required to fulfill the market and industry needs. Consequently, the use of both robust and fragile watermarking techniques for HEVC video standard is very helpful, which requires a huge concern of researchers. Hence, watermarking techniques can be mainly used for:

\textbf{Authentication:} The wide range of video applications, the high accessibility to video, and the availability of video editing software allow unauthorized personnel to easily tamper any video, which highly concerns many video production companies \cite{1}. Therefore, this concern has motivated the need for authentication functionalities to detect and localize any unauthorized tampering in HEVC videos, where fragile or semi-fragile watermarks could be used in order to do so. These types of watermarks are aimed to be destroyed in case of any unauthorized alteration of watermarked videos.

\textbf{Copyright protection:} It concerns the identification of content owners to protect their ownership. Thus, robust watermarks have to be used for copyright protection to ensure that watermarks are persistently associated with the video contents. Robust techniques rely on watermarking stable zones in the video to ensure that the copyright information is all the time existing in the video to identify its owner.

\subsection{Q3: What are the current watermarking techniques that are used or can be applied for the HEVC codec to provide authentication and copyright functionalities?} \label{classification}
\begin{figure*}[!h]
	\centering
	\includegraphics[width=0.8\linewidth]{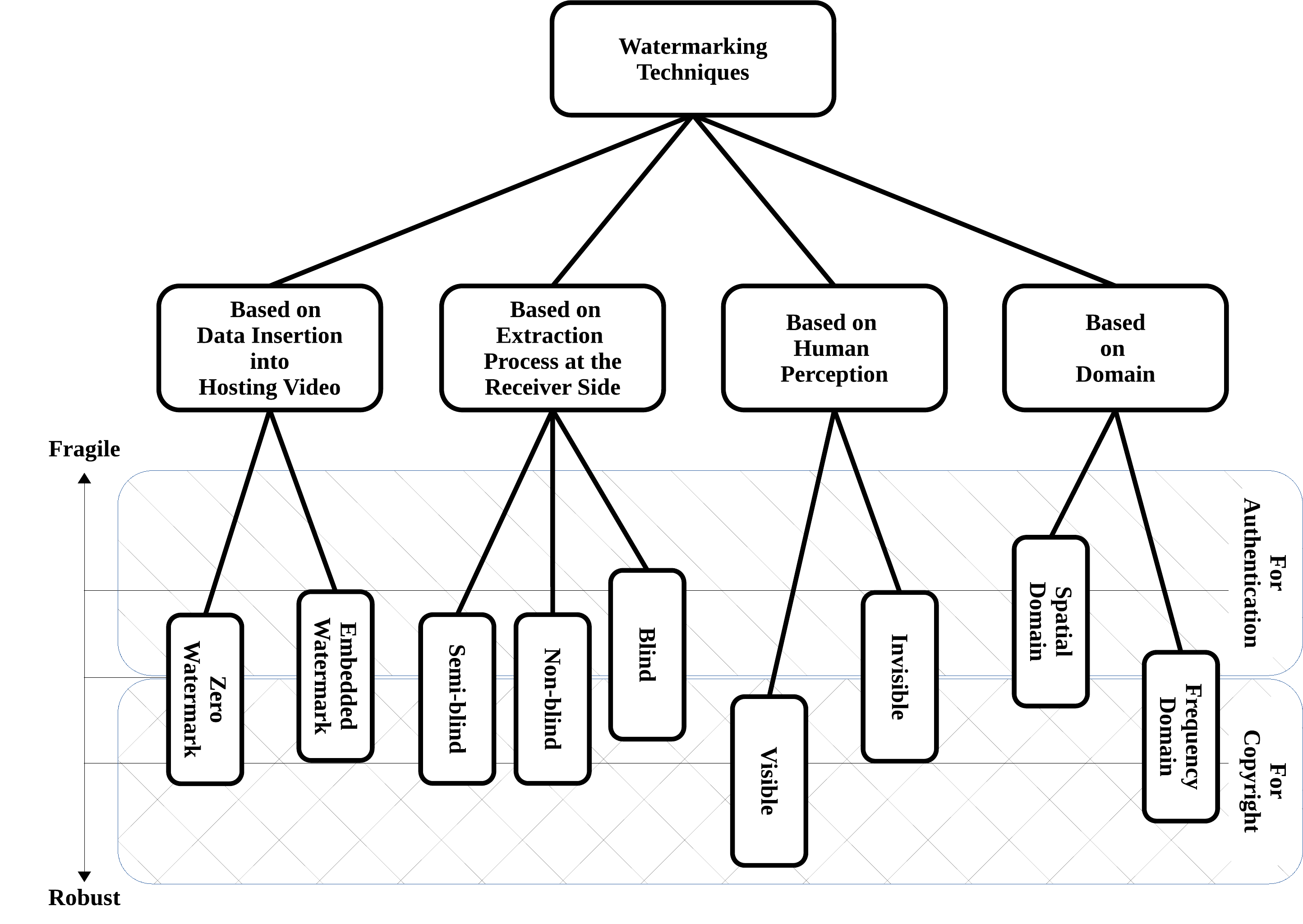}
	\caption{Classification of the commonly used watermarking techniques.}
	\label{fig:fig2}
\end{figure*}

Video watermarking has different techniques depending on the targeted application; fragile/semi-fragile techniques are commonly used for authentication purpose and robust techniques are used for copyright applications. In general, the watermarking methods designed for the previous standards; such as MJPEG, H.263, and H.264; can be applied to the HEVC. However, these watermarking techniques cannot be straightforwardly applied to the HEVC standard due to its new features and tools, which have not been considered in the previous codecs. Thus, the existing watermarking techniques have to be improved or modified to fit the environment of the HEVC standard and its requirements. These watermarking techniques can be categorized as in Fig. \ref{fig:fig2}, which is a combination of new information with some common classifications presented in previous literature \cite{20, 21, 22, 23}. 

\subsubsection{\textbf{Based on the domain:}}
The domain is a place or a stage in which the watermark is embedded, which can be categorized into two main categories:
\begin{enumerate}
	\item[1.1] \textit{\textbf{Spatial domain:}} in this domain, the watermarking can be performed in a bitstream-wise or in a pixel-wise manner, as explained below:
		\begin{itemize}
			\item \textit{\textbf{Bitstream-wise:}} the watermarking is done at the entropy part of the codec, where the watermarking can be done using either the Least Significant Bit (LSB) or the Most Significant Bit (MSB). More specifically, the watermark is directly embedded into or extracted from a compressed bitstream, which can be embedded as a string of bits in the Context Adaptive Binary Arithmetic Coding (CABAC) of the HEVC entropy part or it can be extracted as a feature from the bitstream itself, similarly as proposed in \cite{54}, in which a string of bits was embedded into the H.264 CABAC. Moreover, watermarks can be embedded into the motion vector data, which is a particular case of bitstream domain technique, where watermarks can be inserted into the MSB in the motion vector data \cite{3, 55}.
			
			Indeed, this approach presents high fragility, which is significantly proper for authentication and inappropriate for copyright applications. The watermarking in this domain is proportionally simple and proper for real-time video streaming since it works directly on the bitstream. However, this approach results in some critical issues including visual distortion and fragility against common image processing operations; such as rotation, resizing and video re-compression, etc.
			
			\item \textit{\textbf{Pixel-wise:}} in this domain, the watermarking is done based on the hosting frame statistical features such as histograms \cite{zong2014, rajkumar2019} and moments \cite{duan2014} that can be used as an extracting-based or embedding-based watermarking technique. More specifically, these features can be used in a non-blind manner, in which the decoder needs information about the hosting frame to validate the watermark. Otherwise, these features can be used in a blind manner, in which the information of the hosting frame is not required at the decoder side. Similar to the bit-wise domain, the pixel-wise domain suffers from high complexity, visual distortion, and high fragility against video re-compression, etc. That is why it can be much more suitable for authentication applications.
		\end{itemize}
	 
	\item[1.2] \textit{\textbf{Frequency domain:}} in this domain, there are many common approaches, such as Discrete Sine Transform (DST) and Discrete Cosine Transform (DCT). Despite that the watermarking techniques in this domain are highly robust, they still suffer from high complexity issues \cite{1,13,36,83,84}. They are commonly used for both copyright and authentication applications due to their high robustness against unintentional attacks. However, they are not proper for real-time video streaming due to their complexity unless parallel processing is considered.
\end{enumerate}

\subsubsection{\textbf{Based on human perception:}}
In this category, the watermarks can be either visible or invisible:
\begin{enumerate}
	\item[2.1] \textit{\textbf{Visible watermarking:}} is mostly used for copyright protection. However, this technique can be highly fragile if the watermark is placed in a non-significant part of the video, where it can be covered, cropped, or removed. In fact, the robustness of this technique could be significantly improved if the visible watermark changes its location randomly over the video time, which makes the watermark removal an uneasy task. Therefore, visible watermarking requires to be perceptible and non-removable. However, it is still not easy for video owners to detect illegal video distribution when visible watermarks are used. For this reason, video owners need to rely on invisible watermarks for better tracking of illegal videos \cite{60}. 	
	
	\item[2.2] \textit{\textbf{Invisible watermarking:}} it can be fragile or robust depending on how and where the watermark is embedded in the video. Therefore, the invisible watermarking techniques are very commonly used for both copyright and authentication applications. According to the resistance level against intentional or unintentional attacks, invisible watermarking techniques can be further classified as robust or fragile \cite{20,62}. Specifically, the watermarking technique is considered robust against attacks if the watermark can survive after the watermarked video is altered or attacked. These robust watermarking techniques are primarily suitable for copyright protection applications. In contrast, fragile watermarking is a technique in which the watermark will be destroyed if the watermarked video is altered or attacked. These fragile watermarking techniques are generally suitable for authentication applications as well as for tampering detection.
\end{enumerate}

\subsubsection{\textbf{Based on the extraction process at the receiver side:}}
In this category, there are three main classes:
\begin{enumerate}
	\item[3.1] \textit{\textbf{Blind watermarking:}} is a way of extracting watermarks without using their original information at the receiver side to detect and validate them, as shown in Fig. \ref{fig:blind}. This category can be visible or invisible and it can be used for copyright protection and authentication. The main problem with this category, when it is being used for copyright, is that the watermark could be non-detectable if the watermarked video encountered some attacks, such as frame dropping and/or filtering. To overcome this problem, the Forward Error Correction (FEC) approach can be used to detect and correct errors in the recovered watermark, as proposed in \cite{13}. Moreover, statistical approaches; such as moments, histograms, and hybrid transforms; can be also used to overcome this issue, as proposed in \cite{76, zong2014, rajkumar2019, duan2014}.
	
	\begin{figure*}[h!]
		\centering
		\includegraphics[width=0.7\linewidth]{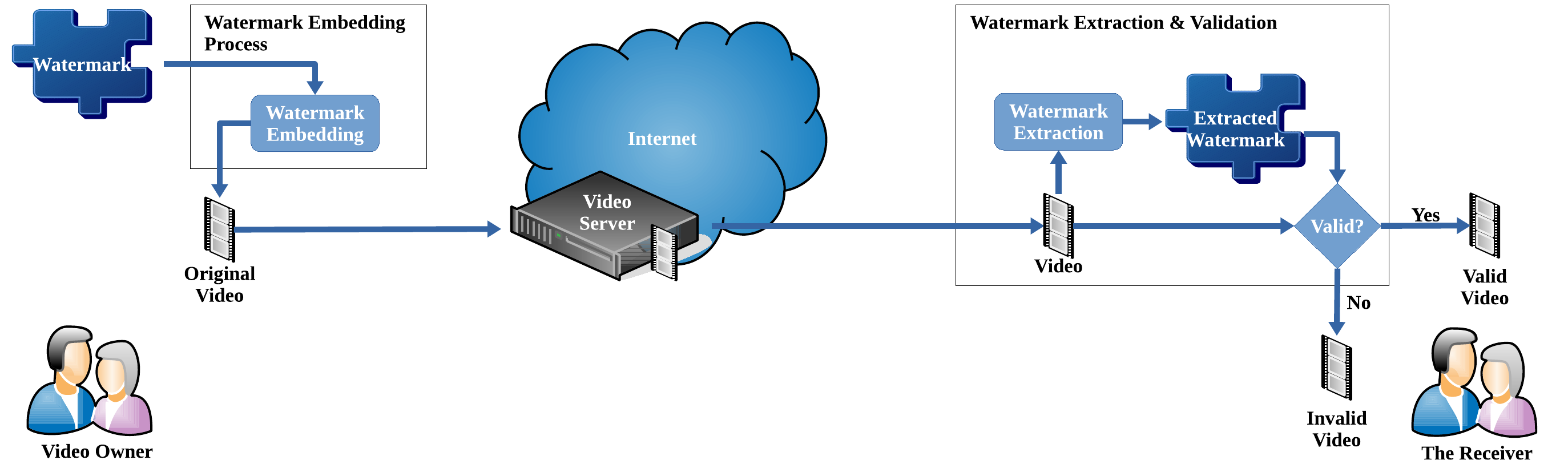}
		\caption{Blind Watermarking Architecture.}
		\label{fig:blind}
	\end{figure*}
	
	\item[3.2] \textit{\textbf{Non-blind watermarking:}} is an approach used for extracting watermarks at the receiver side using the original watermark information to detect and validate them, as shown in Fig. \ref{fig:nonblind}. This type can be visible or invisible and it is appropriate for both authentication and copyright protection. The most common approach of this category is the zero-watermarking, which relies on the features of the video itself that need to be known at the receiver side in order to be able detectable and validatable \cite{27, 49, 56, 57, 63}. The main issue with this type is that an intermediate authentication server is required for the transmitter to share the original watermark information with the receiver to use it for watermark extraction and validation, which increases complexity and cost of the implementation of this approach.

	\begin{figure*}[h!]
		\centering
		\includegraphics[width=0.7\linewidth]{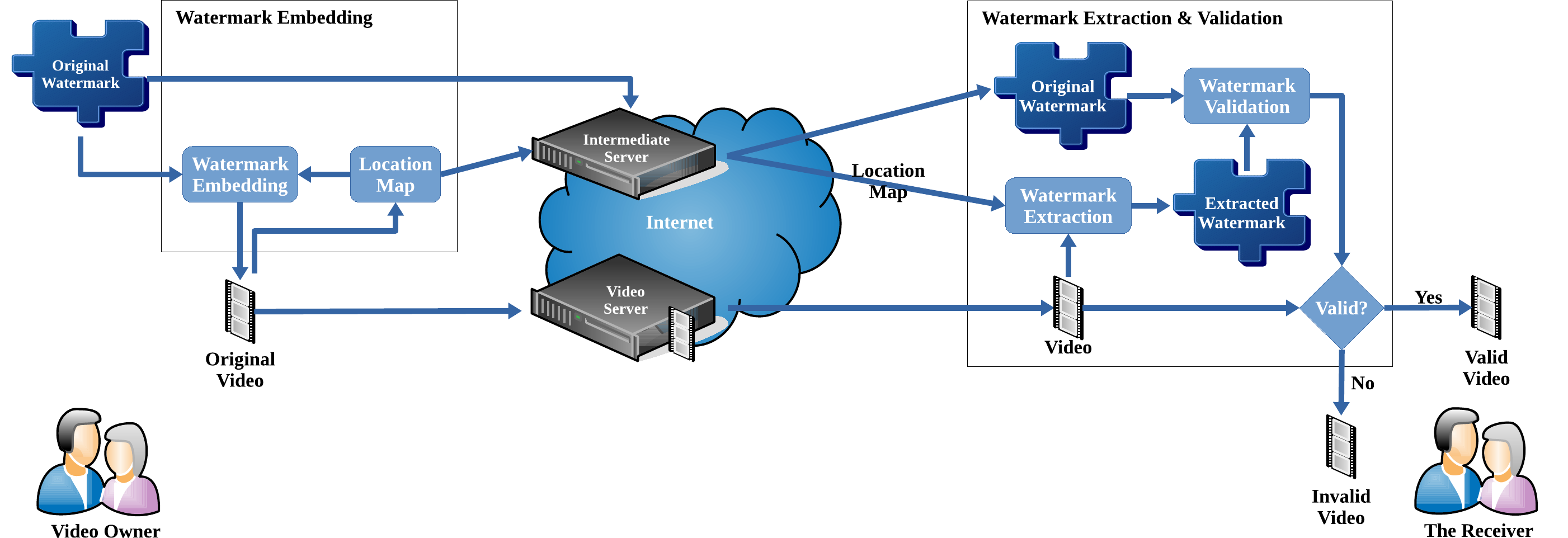}
		\caption{Non-Blind Watermarking Architecture.}
		\label{fig:nonblind}
	\end{figure*}
	
	\item[3.3] \textit{\textbf{Semi-blind watermarking:}} is an approach used for extracting watermarks at the receiver side using the location map that points to the used positions to embed the watermark into the hosting video, in order to make the watermark detectable and validatable, as shown in Fig. \ref{fig:semiblind}. This type can be visible or invisible and it is appropriate for both authentication and copyright protection. Even though this approach helps to reduce the sensitivity to synchronization errors, it is still suffering from some critical issues such as the extra transmission overhead due to attaching the location map with the watermarked video \cite{12}. In order for researchers to avoid this issue, a secure channel through an intermediate authorization server is used to send the location map to the receiver, which inherits the same issue introduced by the non-blind method \cite{8, 70, 85, 98, 99}.
	
	\begin{figure*}[h!]
		\centering
		\includegraphics[width=0.7\linewidth]{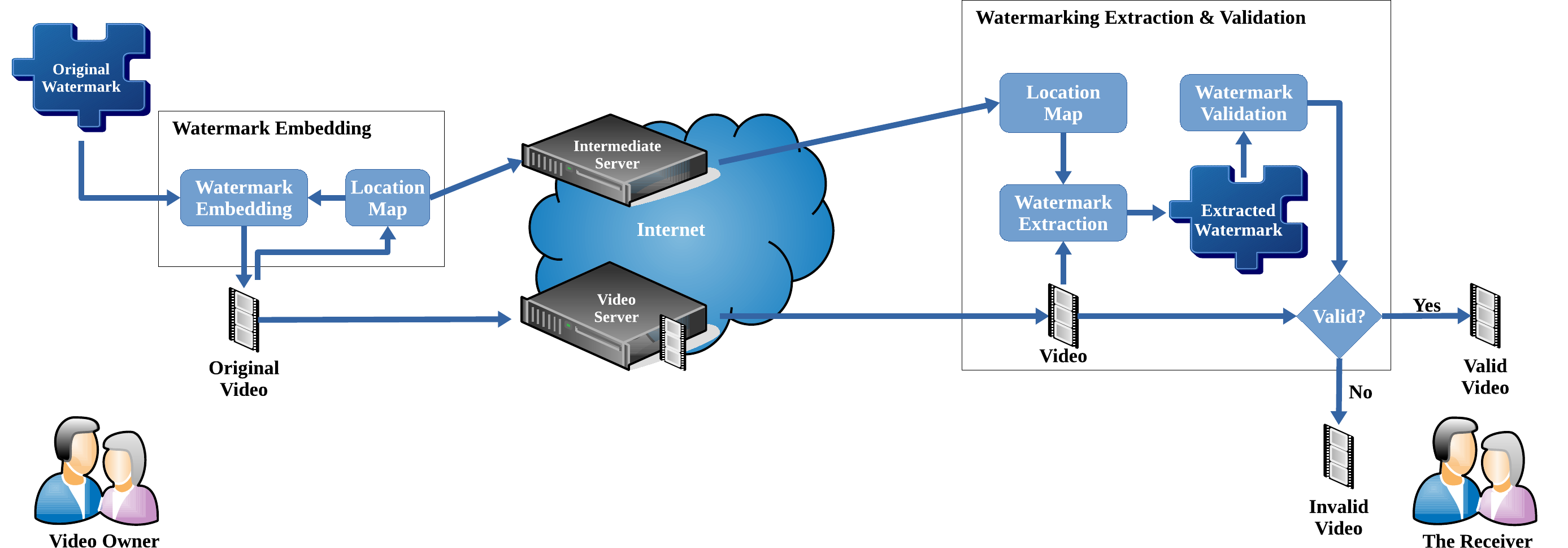}
		\caption{Semi-Blind Watermarking Architecture.}
		\label{fig:semiblind}
	\end{figure*}
\end{enumerate}

\subsubsection{\textbf{Based on the data insertion into hosting video:}}
In this category, there are two main classes:
\begin{enumerate}
	\item[4.1] \textit{\textbf{Zero-watermarking:}} can be used for authentication and copyright protection applications. It works based on extracting some features of the video itself to be used as a watermark, thus no information will be embedded into the original video, as shown in Fig. \ref{fig:zerowm}. These features have to be known at the receiver side in order to make the watermark detectable and validatable, which makes this category falls under the non-blind approach \cite{27, 49, 56, 57, 63}.  
	
	\begin{figure*}[h!]
		\centering
		\includegraphics[width=0.7\linewidth]{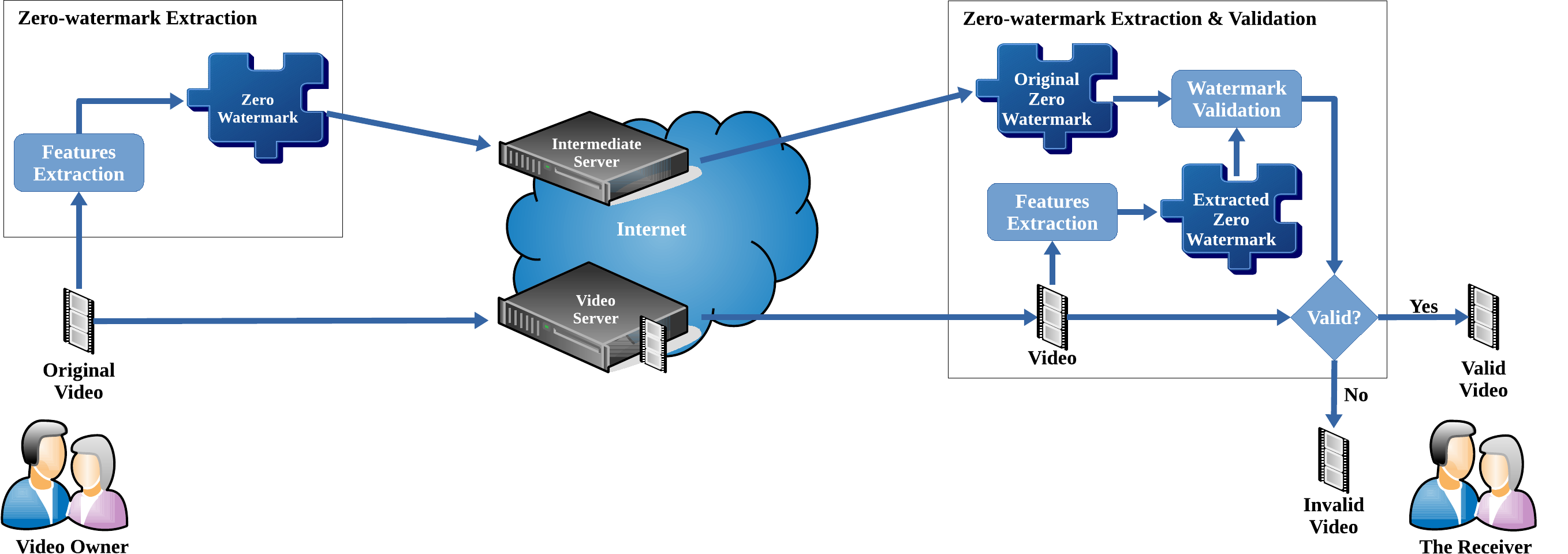}
		\caption{Zero-Watermarking Architecture.}
		\label{fig:zerowm}
	\end{figure*}
	
	Indeed, the zero-watermarking is an excellent solution to avoid the visual quality degradation encountered by the conventional watermarking techniques, where the later can be highly robust on the expense of the video visual quality and bitrate \cite{100}. However, the zero-watermarking approach mainly relies on an intermediate server to share the original watermark information with the receiver to use it for watermark extraction and validation. Specifically, the sender will share the original zero-watermark with the receiver trough out the intermediate server if the used approach is non-blind, while the location map of the extracted features only will be shared with the receiver if the used approach is semi-blind. In general, the intermediate server used in this approach is considered the main weakness that increases its complexity and cost of implementation.
				
	\item[4.2] \textit{\textbf{Embedded-watermarking:}} is a method to inject some external data into a video, where this method can rely on blind approach (as in Fig. \ref{fig:blind}), non-blind approach (as in Fig. \ref{fig:nonblind}) or semi-blind approach (as in Fig. \ref{fig:semiblind}). Embedded-watermarking is suitable for both authentication and copyright protection with less complexity compared to the zero-watermarking techniques. However, this approach is still suffering from some critical problems caused by injecting data into the video, such as (1) video bitrate increase (2) visual quality degradation (3) survivability against the common attacks \cite{1, 8, 11, 12, 70, 85, 98, 99}.
\end{enumerate}

\subsection{Q4: What are the possible options to implement watermarking techniques into the HEVC codec for authentication and copyright applications?}
In general, HEVC watermarks could be embedded into videos (embedding-based watermarking technique) or extracted from videos (zero-watermarking technique). Regardless of the used approach, there are four possible options that can be used to perform video watermarking at the codec level. In Fig. \ref{fig:fig3}, we have highlighted the possible watermarking options for the HEVC video standard:
\begin{figure*}[h]
\centering
\includegraphics[width=0.7\linewidth]{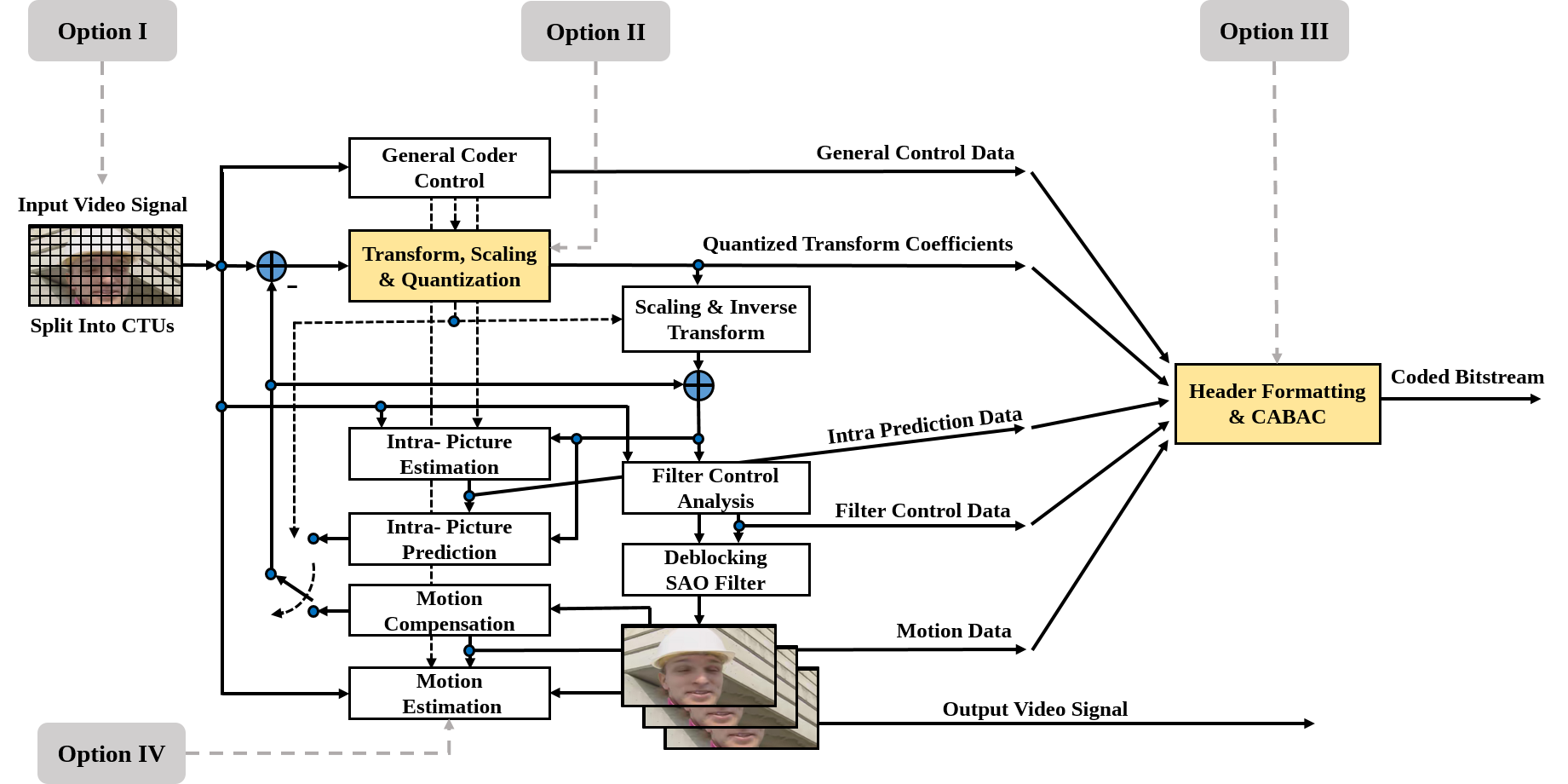}
\caption{Possible Watermarking Options for HEVC Video Standard.}
\label{fig:fig3}
\end{figure*}

\textbf{Option I:} Although the option of using watermarking methods based on the spatial domain in a pixel-wise manner is not new, however, they have only been used in the image watermarking area for many years \cite{zong2014, duan2014, rajkumar2019}. This option is also suitable for video watermarking since videos are chains of sequential images. This option is highly fragile against the common attacks, that is why it can be much more suitable for authentication applications. In this option, the watermarking can be performed based on the hosting frame statistical features, such as histograms and moments that can be used as extracting-based or embedding-based watermarking techniques. However, using histograms and moments could significantly increase the codec complexity and may lead to visual distortion. In general, this option can not be directly applied to the HEVC standard due to some differences between the HEVC and image codecs, such as the variability of the HEVC transform block size that varies from 4x4 to 32x32, and the high embedding capacity of the videos compared to images due to the presence of the time domain in the video. Thus, these differences have to be considered in order to design a successful watermarking technique for the HEVC standard based on the spatial domain in a pixel-wise manner. To the best of our knowledge, there is no HEVC watermarking technique performed based on this option yet. 

\textbf{Option II:} Watermarking can be performed based on the frequency domain techniques as proposed in \cite{4, 73,78} that was designed for H.264. However, these techniques cannot be directly applied to the HEVC standard due to the variability of the transform block size, which varies from 4x4 to 32x32 and also due to the increase of the intra-modes up to 35 in HEVC rather than 9 intra-modes as in H.264 standard. Additionally, the HEVC uses the DST transform for blocks of 4x4 size and the DCT transform for blocks of 8x8 to 32x32 size, while only DCT transform is used in the H.264 standard. Moreover, both H.264 and H.265 rely only on the Y channel for watermarking while the chromatic channel is never used in the literature, especially in the embedding-based watermarking techniques \cite{70, 87}. Thus, these differences have to be considered to design a new watermarking technique fitting the HEVC standard.

Recently, \cite{8,12,13,15,17,38,52} developed new robust HEVC watermarking techniques based on the transform domain that are slightly robust compared to the spatial domain watermarking techniques as in Option I. In \cite{35}, the watermark is embedded into the quantized transform coefficients during the encoding process. However, the authors did not report the robustness level against the common types of attacks. In \cite{16}, a new hiding data method for HEVC based on the intra-frame error propagation effect technique was proposed, in which the blocks are classified based on the intra-prediction modes. However, this method is fragile and very sensitive to common attacks such as image re-compression and image processing attacks \cite{12}.

\textbf{Option III:} The watermark can be directly embedded into the compressed encoded bitstream at the bit-wise spatial domain. However, the H.264 methods used previously in this domain cannot be directly applied to the current HEVC standard, due to differences between the HEVC CABAC and the H.264 CABAC. 

Recently, references \cite{43,47} proposed two hardware-based HEVC watermarking techniques for ownership verification. Both techniques use AC-NZ coefficients into the arithmetic algorithms at the CABAC for embedding and extracting watermarks, where they also control the bitrate increase and minimize the complexity. However, both techniques suffer from slightly high drift and synchronization errors due to the CABAC open loop that lacks feedback to minimize errors. Moreover, using this option is fragile to some extent and sensitive to common attacks such as re-compression and image processing attacks. For more details, refer to Q3-1.1.

\textbf{Option IV:} Watermarking can be also performed on motion vector data, which can be defined as a particular case of bitstream domain technique \cite{3, 55}. The watermark can be inserted into the MSBs of motion vectors. However, to apply the watermarking on motion vector data in HEVC should consider the differences between the motion-compensated prediction algorithm applied in the HEVC codec and the motion-compensated prediction algorithm applied in the H.264. To the best of our knowledge, there is no HEVC watermarking technique performed using this option yet.

\subsection{Q5: What are the main watermark zone selection criteria that can be applied to design efficient HEVC watermarking techniques for authentication and copyright applications?}
The watermark zone selection is very crucial for the embedding process, the extraction process, and the targeted application. For instance, the main goal of watermarking techniques, targeting copyright applications, is to survive against attacks in order to protect intellectual property rights. Therefore, the watermarking techniques, in this case, would select the invariant or stable regions of the video content to embed or extract the copyright information, which in turn would increase the robustness of the technique. On the other hand, the watermarking techniques, targeting authentication applications, would select the variant or unstable regions of the video content in order to increase the sensitivity to any attack, which is used to detect and localize any tampering on the video content. In general, the watermark zone selection can be done based on four main criteria, as described below:

\subsubsection{\textbf{Color channel selection:}} The watermarking techniques use Luminance channel (Y) for embedding watermark information because the modification in the Y channel would not affect the Human Visual System (HVS) compared to embedding watermark information in chromatic channels. Moreover, the Y channel cannot be removed without causing a significant impact or total damage of the watermarked video, while the chromatic channels can be removed without affecting the visual quality. More specifically, removing the chromatic channels of the video will not cause a visual distortion but will remove colors from the video and will convert it to the grayscale mode \cite{70, 87}. 

\subsubsection{\textbf{Frame selection:}} Generally, video codecs divide the video stream into Group of Pictures (GOP), which includes three types of frames; I, P, and B. The I-frame is an intra-coded frame, which is coded independently of all other frame types. As regarding the P-frame, it is a predictive coded frame that contains the motion-compensated difference information relative to the previously decoded frame, while the B-frame is a bi-predictive coded frame that contains the motion-compensated difference information relative to the previously decoded frames. In fact, the P and B frames are considered as unstable and sensitive areas due to containing a small Number of Non-Zero (NNZ) coefficients. For this reason, both could be used to embed/extract watermarks designed for highly fragile authentication applications. However, using them for watermarking will lead to visual quality distortion, bitrate increase, and increase of vulnerability to the common attacks. As for the I-frames, they are classified as stable areas due to containing large NNZ coefficients that can be used for embedding and/or extracting watermarks if a robust watermarking technique is needed. Additionally, I-frames can be used for watermarking techniques designed for authentication purposes as well \cite{1,67}.

\subsubsection{\textbf{Block selection:}} The blocks in each frame are classified into stable blocks and unstable blocks based on the following parameters: the size of the block, NNZ coefficients, and motion information. 

\begin{enumerate}
	\item[3.1] \textit{\textbf{Block size:}} The intra-prediction modes of HEVC have block sizes of 32x32 down to 4x4. In each frame, a block is considered stable if its size is equal to 4x4, otherwise, it would be considered as an unstable block, where the unstable blocks have limited details compared to stable blocks that are very rich of details. For this reason, the unstable blocks are prone to be changed into other block sizes if they were exposed to any attack process \cite{12,34}. To summarize, 4x4 blocks could be used for watermark embedding or feature extraction regardless the type of the frame (I, P or B), however, 4x4 blocks in the I-frame are preferable for robust watermarking techniques while 4x4 blocks in P-/B-frames are much suitable for fragile techniques that could be used for authentication applications \cite{1,8,12}.
	
	\item[3.2] \textit{\textbf{The NNZ quantized coefficients:}} It plays a significant role during the block selection process, where it can be used to differentiate between stable blocks and the extremely stable ones. Specifically, 4x4 blocks with high NNZ coefficients contain a lot of picture details compared to other blocks with the same size with less NNZ. Thus, blocks with high NNZ coefficients have a high level of stability that could be utilized for watermark embedding or feature extraction. Moreover, the use of blocks with high NNZ coefficients for watermarking will lead to trivial bitrate increase and less sensitivity to synchronization errors \cite{12,34,70}.
	
	\item[3.3] \textit{\textbf{Motion information:}} It plays another critical role in choosing stable blocks for watermarks embedding or feature extraction that can reduce the impact of synchronization error. As well-known, the frames could contain static, low-motion, or high-motion picture components, where frames with low-motion components are considered highly stable with low sensitivity to synchronization errors \cite{12}.
\end{enumerate}

\subsubsection{\textbf{Coefficient selection:}} It is an essential part to minimize the propagation drift and the bitrate increase, and also to minimize the impact of the synchronization errors. For instance, the blocks can be classified into groups based on the intra-prediction modes, to determine the protected pixel sets regardless of their block sizes to be used for embedding without causing the propagation drift error \cite{16}. Similarly, Reference \cite{38} applied the same approach proposed in \cite{16} but only uses blocks of 4x4 size to produce a robust watermarking technique with minimal propagation drift errors. 

Since using the zero-frequency DC coefficients for watermark embedding will increase the bitrate and the visual distortion, AC coefficients are considered more suitable to reduce the aforementioned problems. Moreover, using the high-frequency AC coefficients will increase the fragility of the watermarking technique because it could be removed intentionally or unintentionally using image processing operations. As for the middle-frequency AC coefficients, they could be the key point to reduce the visual distortion errors due to the difficulty of recognition by the human eyes compared to the DC coefficients. Furthermore, using the middle-frequency AC coefficients for watermark embedding could highly increase robustness and resistance of the technique against the synchronization errors and the image processing operations \cite{70}. 

\subsection{Q6: What are the common metrics used to evaluate the performance of video watermarking techniques?}
In order to evaluate a watermarking technique, there are several evaluation metrics that could be used depending on the target of the technique.

\subsubsection{\textbf{Peak signal to noise ratio (PSNR):}} PSNR is commonly used to examine the quality of the image or frame, which is calculated as the ratio between the maximum possible power of the original image and the power of the watermarked image. According to \cite{61}, the PSNR for YUV video sequence is calculated using Equation \eqref{eq1} as follows:
\begin{align}
	\label{eq1}	
	PSNR = \frac{(6 \times PSNR_Y) + PSNR_U + PSNR_V}{8},
\end{align}
where $PSNR_Y$ is the PSNR of the Luma component, and $PSNR_U$ and $PSNR_V$ denote the color components. 

\subsubsection{\textbf{Bit error rate (BER) and watermarking robustness rate (WRR):}} BER and WRR are used to measure the robustness of watermarking techniques. According to \cite{12, 86}, the BER and WRR are defined as in the equations \eqref{eq2} and \eqref{eq3}, respectively:
\begin{align}
\label{eq2}	
 BER=\frac{E_b}{T_b},
\end{align}
where $E_b$ is the number of error bits and $T_b$ is the total bits sent, while WRR is the complement of the BER calculated as below:
\begin{align}
\label{eq3}	
WRR=1-BER
\end{align}

\subsubsection{\textbf{Normalized cross-correlation (NCC):}} NCC is also used to measure the robustness of watermarking techniques, by determining the similarity between the original watermark $W$ and the recovered watermark $W^{'}$, as computed in Equation \eqref{eq4} \cite{75} below:
\begin{align}
	\label{eq4}	
	NCC=\frac{\sum\limits_{m=1}^{M}\sum\limits_{n=1}^{N}W(m,n)W^{'}(m,n)}
	{\sqrt{\sum\limits_{m=1}^{M}\sum\limits_{n=1}^{N}W(m,n)^{2}\sum\limits_{m=1}^{M}\sum\limits_{n=1}^{N}W^{'}(m,n)^{2}}},
\end{align}
where $m$ and $n$ are the dimensions of the watermark.

\subsubsection{\textbf{Bit increase rate (BIR):}} BIR is another important metric used to measure the efficiency of a video watermarking technique by computing the bitrate increase after embedding a watermark in a hosting video. The bitrate can be calculated as $Bitrate (bits/sec)=\frac{V}{T}$, where $V$ is the video file size in bits and $T$ is the playback time in secs. As for the percentage of increase, it can be calculated as in Equation \eqref{eq6} \cite{40,85} below:
\begin{align}
	\label{eq6}	
	BIR =\frac{WV_{br} - O_{br}}{O_{br}}\times 100,
\end{align}
where $WV_{br}$ represents the watermarked video bitrate and $O_{br}$ denotes the original video bitratre.

\subsubsection{\textbf{Embedding capacity ratio per frame (ECRF):}} ECRF is used to measure the embedding capacity of the watermarking technique, which can be calculated by dividing the amount of embedded watermark data on the total amount of cover frame, as in Equation \eqref{eq7} \cite{70}:
\begin{align}
\label{eq7}	
ECRF =\frac{D_{embedded}}{F_{size}}\times 100 ,
\end{align}
where $D_{embedded}$ is the size of embedded data and $F_{size}$ is the size of the hosting frame before embedding the watermark data, where the higher the ECRF is the better.

\subsection{Q7: What are the main challenges in HEVC video watermarking?}
Indeed, most HEVC watermarking techniques are suffering from many challenges; where some of these challenges are common and especially affecting watermarking techniques designed for both authentication and copyright applications. Moreover, some of these challenges could affect the watermarking techniques depends on the application, where some could affect the techniques designed for authentication while some could affect techniques designed for copyright.

\subsubsection{\textbf{Common challenges:}}
The common challenges discussed in this subsection could affect the HEVC video itself and/or the HEVC bitstream. Specifically, under this category of challenges, we have two main issues as follows:

\paragraph{Visual distortion issue:}
In fact, visual distortion could happen only when embedding-based watermarking techniques are used, while it could never happen when using zero-watermarking techniques. Generally, there are two main causes of visual distortion when embedding-based watermarking techniques are used with HEVC:
\begin{itemize}
	\item The use of AC-Zero or DC coefficients has a severe negative impact on the visual quality of the HEVC video, due to the increase of errors during the reversing procedure of the video at the decoder side. Moreover, the change of the aforementioned coefficients will also lead to a bitrate increase of the HEVC video, which is not a preferable scenario. For this reason, the watermark information has to be embedded into the AC Non-Zero (NZ) coefficients to reduce potential errors during the HEVC video retrieval process at the decoder side.
	
	However, using the AC-NZ coefficients alone still cannot guarantee the avoidance of video visual distortion. For example, references \cite{1,8,11,14,17,26,30} proposed embedding-based watermarking techniques based on using AC-NZ coefficients to avoid the visual distortion but all of these proposed techniques suffered from the error drift issues that are still affecting the visual quality of watermarked videos while the bitrate is maintained at the minimum level.

	\item The error drift is defined as an error caused by modifying some pixels of a block that affect or change the pixels of other blocks. Specifically, this error propagates to several blocks in the same frame or other frames predicted from the modified block, which yields degrading in the visual quality of the watermarked video. Therefore, to avoid this problem, the selection of coefficients is significantly important to protect the adjacent blocks from being affected by the propagated error(s) caused by modifying a certain block.
	
	The first research to prevent the error drift issue on HEVC videos was proposed by Chang et al (2014) \cite{16}. It classified pixels in two (protected and unprotected) pixels groups out of all block sizes, where the coefficients generated from the protected pixels group is used for embedding information to avoid intra-coded frame errors. However, this technique suffers from minor visual distortion due to the non-precise selection of the protected pixels group. 
	
	Reference \cite{38} finds the residual blocks by applying the inverse discrete sine transform on the 4x4 blocks only to determine the protected set of pixels to find a 3x3 residual protected subblock out of the selected residual 4x4 block. This approach showed an acceptable level of robustness but it could not completely prevent the distortion drift due to quantization errors that propagate to the neighboring blocks \cite{44}.
	
	Reference \cite{44} embeds the watermark into the intra-prediction residual 4x4 blocks at the spatial domain. This approach showed an acceptable level of robustness while the distortion drift errors were significantly avoided. However, this technique is not robust enough against the frame dropping/exchange attack, for this reason, it is highly recommended to use the error correction codes to strengthen it against such attack.

	In reference \cite{83}, 4x4 intra-luma transform blocks of I-frames are divided into two sets: (1) Robust set is used for embedding the generated code in order to make this code survival against unintentional attacks; such as re-compression attacks, noise attacks, and frame dropping attacks; while (2) Fragile set is used to generate authentication code in order to check the video integrity. Despite that it works fine for both copyright and authentication applications, however, it showed a slight visual distortion due to the un-consideration of drift errors. Recently, \cite{92} proposed a new technique based on a matrix encoding approach that maintains free error drift with high embedding capacity, high fidelity, and minimum visual distortion.
\end{itemize}

\paragraph{Bitrate increase issue:}
As mentioned early, watermarking techniques should straightforwardly use the AC-NZ coefficients to minimize the bitrate increase, especially when constrained channel capacity is used \cite{24} as proposed in \cite{1,12,13,16,35,37,94}. As well-known, the NNZ coefficients are directly proportional to the nature of the video itself in terms of richness of details and motion. Therefore, videos with low details and motion could have low NNZ coefficients that could be insufficient to carry the watermark information. Consequently, this may lead to the use of zero or DC coefficients which will severely affect the visual distortion and the bitrate increase.

\paragraph{Synchronization error issue:}
This issue is very important since it negatively affects any watermarking techniques regardless of what approach is used. As aforementioned in Q5, embedding a watermark into an HEVC video using inaccurate selection criteria could cause drifting errors that could lead to synchronization issues. However, synchronization errors also happen due to the size change of blocks located in smooth regions of any HEVC video frames if the video is exposed to any attack, which leads to extraction failure of the watermark at the decoder side \cite{12,13}. 

In fact, using accurate selection criteria to find the appropriate blocks for watermarking is a crucial task to reduce the probability of synchronization error, however, it can not be completely avoided. For this reason, implementing an error detection and correction code is highly recommended for watermarking techniques to minimize the synchronization errors \cite{13}.

\subsubsection{\textbf{Authentication challenges:}}
Video watermarking is considered efficient for authentication if it could detect and localize any tampering on video content. Usually, fragile watermarking techniques are used for authentication purposes, in which they could employ either embedding-based watermarking or zero-watermarking techniques.

However, HEVC video watermarking for authentication purposes still faces many challenges. Therefore, it is essential to highlight the arisen issues in the study of applying fragile watermarking techniques on HEVC standard to make it easier for researchers in this field to have a comprehensive overview of the most common issues and challenges. For better understanding, we listed these issues and challenges in this subsection with a brief description, discussion, and analyses. Additionally, Table \ref{tbl:tbl3} shows a brief comparison of the latest HEVC watermarking techniques based on their authentication capabilities.

\begin{table}[h]
\caption{Comparison of the latest HEVC video watermarking techniques based on their authentication capabilities} \label{tbl:tbl3}
	\begin{center}\scriptsize
		\begin{tabular}{cllp{1cm}p{1cm}p{1cm}}\hline
			\multicolumn{1}{l}{\multirow{4}{*}{Year}}&\multicolumn{1}{l}{\multirow{4}{*}{Ref\#}}&\multicolumn{1}{l}{\multirow{4}{*}{}}&\multicolumn{3}{c}{Capabilities}\\\cline{4-6}
			 	 						& 			& Watermarking		& Tampering & Localization  & Double			 \\
				 						& 			& Technique			& Detection & Ability		& Compression		 \\
				 						&			&					& 			&				& Detection			 \\\hline
			\rowcolor[HTML]{EFEFEF}2015 & \cite{11} & Embedding-based	& Yes 		& Yes			& No		 		 \\
								   2015 & \cite{26} & Zero-watermark  	& No   		& No			& Yes*		 		 \\	
			\rowcolor[HTML]{EFEFEF}2016 & \cite{14} & Embedding-based 	& Yes 		& Yes			& No		 		 \\	
								   2017 & \cite{7}  & Zero-watermark 	& Yes 		& No			& Yes		 		 \\
			\rowcolor[HTML]{EFEFEF}2018 & \cite{1}  & Embedding-based	& Yes  		& Yes   		& No		 		 \\			
								   2018 & \cite{27} & Zero-watermark 	& Yes 		& NM			& NM		 		 \\
			\rowcolor[HTML]{EFEFEF}2018 & \cite{45} & Zero-watermark 	& Yes		& No			& Yes		 		 \\
								   2018 & \cite{46} & Embedding-based 	& No		& No			& Yes*		 		 \\
			\rowcolor[HTML]{EFEFEF}2018 & \cite{95} & Zero-watermark 	& No		& No			& Yes		 		 \\
								   2019 & \cite{50} & Zero-watermark 	& No		& No			& Yes*		 		 \\
			\rowcolor[HTML]{EFEFEF}2019 & \cite{53} & Zero-watermark 	& No		& No			& Yes		 		 \\
								   2019 & \cite{83} & Embedding-based 	& Yes		& Yes			& No		 		 \\
			\rowcolor[HTML]{EFEFEF}2019 & \cite{84} & Zero-watermark 	& No		& No			& Yes*		 		 \\ 
								   2019 & \cite{89} & Zero-watermark 	& No		& No			& Yes		 		 \\ 
			\rowcolor[HTML]{EFEFEF}2020 & \cite{93} & Embedding-based	& Yes		& Yes			& No		 		 \\
								   2020 & \cite{91} & Zero-watermark	& No		& No			& Yes		 		 \\\hline
			\multicolumn{6}{l}{(*) = Works well only with different bitrates}\\
			\multicolumn{6}{l}{(NM) = Not Mentioned, (NA) = Not Applicable}
		\end{tabular}
	\end{center}
\end{table}

Tew et al. (in \cite{1, 11, 14}) proposed three authentication watermarking techniques that embed some extracted features from the video and its encoding parameters. These embedded features are used at the decoder side to verify the integrity and to localize the tampering in the video. In 2015 \cite{11}, they proposed the first authentication technique for the HEVC standard with two layers, detection layer, and localization layer. It works based on generating the authentication code as a sequence of pairs, where each pair is a combination of the position and size of each used block in order. In 2018 \cite{1,14}, they improved their previous work by involving an encryption technique as a new layer, namely the verification layer. However, the proposed techniques in \cite{1, 11, 14} introduced major visual distortion and bitrate increase caused by error drift propagation issues due to their dependency on fragile embedding-based watermarking techniques. Moreover, these techniques were not able to distinguish between the re-compression attack and other types of attacks.

In 2018, Jiang et al. \cite{27} registered the first patent of using a zero-watermarking technique for HEVC video streaming against the re-compression attack. It directly generates a zero-watermark based on processing depth features of the video stream without any modification on the original video. Thereafter, the watermark is encrypted and registered in the third-party institutions. Despite the disclosure of the way of detecting the re-compression attack with different quantization parameters, the re-compression attack with the same quantization parameters was not mentioned in this patent.

In 2019 \cite{83}, the authors divided all 4x4 intra-luma transform blocks of I-frames into two sets: (1) Fragile set is used to generate authentication code in order to check the video integrity; while (2) Robust set is used for embedding the generated code in order to make this code survival against unintentional attacks; such as re-compression attacks, noise attacks, and frame dropping attacks. Even though it works fine for both copyright and authentication applications, however, it showed a slight visual distortion due to neglecting the issue of drift errors.

Indeed, the detection of double compression in HEVC is a significant feature in the authentication domain since most of the common video tampering attacks are usually applied to the spatial domain. Thus, the attacker decompresses the sequential video stream into frames to get ready for video tampering or modification, then, the tampered video has to be re-compressed and injected again as a video stream in the channel. Hence, double compression is still an essential issue in the video authentication process \cite{25}.

In 2015, Huang et al. \cite{26,46,84} presented a technique for distinguishing double compression attacks by analyzing the distribution of NNZ coefficients for the smooth regions of video frames. Specifically, it calculates the difference between the frequency of residual coefficients of the first and second compressions. Unfortunately, this technique failed to detect the second compression due to negligible frequency when the same quantization parameters were used in both compressions.

In 2017 \cite{7}, a new detection technique for double HEVC compression with the same quantization parameters was proposed based on some features extracted from the 4x4 transform blocks of the I-frames. The technique was able to detect double compression attacks even if the same quantization parameters are used. However, Jiang et al. 2019 \cite{89} proposed a similar technique that relies on features extracted from different block sizes instead of 4x4 only, which could achieve better results compared to \cite {7}. 

Moreover, Yu et al. (2020) \cite{91} proposed a technique for distinguishing double HEVC compression by mixing the two approaches proposed in \cite{7, 89}. However, this work detects only whether the HEVC videos were double-compressed by a third party or not, even if it is re-compressed by the same bitrate. Unfortunately, this work has not mentioned the tampering localization issue and the detection of other intentional and unintentional attacks was neglected as well. 

References \cite{45,53,95} proposed three algorithms based on Prediction Unit (PU) features that are used to distinguish between the first and second compressions. The main shortage of these three algorithms was neglecting all other features such as transform blocks properties and coefficients distributions, which led to lower accuracy compared to \cite{7, 50, 89}. 

Recently in 2020, reference \cite{93} studied the selection of relevant regions in I-frames under predefined restrictions and conditions based on transmission specifications. It uses the stable regions to embed the invariant features and the unstable zones of the I-frames to embed a fragile watermark. Simulation results showed high resistance to some common attacks, however, the embedding capacity of the proposed algorithm was very limited, which significantly affects the recovery quality of the watermark and the PSNR value if the size of the embedded watermark increases. 
  
\subsubsection{\textbf{Copyright challenges:}}
Video watermarking techniques for copyright applications are required to be as robust as possible, where the robustness term implies the capability of a watermark to resist the common attacks in order to reach safely to the decoder side. Usually, robust watermarking techniques are used for copyright purposes, in which they could employ either embedding-based watermarking or zero-watermarking techniques \cite{17, 28, 29, 30, 31, 32, 33, 34}. 

In the meantime, the state-of-the-art literature on HEVC watermarking for copyright is still in its initial stage. In this SLR, we found a limited number of researches on HEVC watermarking techniques for copyright applications. Therefore, it is very important to highlight the arisen issues in the area of applying robust watermark techniques to the HEVC standard for this category of applications. In the upcoming paragraphs, we present a brief description, discussion, and analysis for the most common issues hindering the robustness measure. Moreover, Table \ref{tbl:tbl4} shows a comparative summary of the current robust HEVC video watermarking techniques based on their resistance against the common attacks.
\begin{table}[h!]
	\caption{Comparison of the latest HEVC video watermarking techniques based on their resistance against attacks} \label{tbl:tbl4}
	\begin{center}\scriptsize{
		\begin{tabular}{lp{0.3cm}p{0.3cm}p{0.3cm}p{0.3cm}p{0.3cm}p{0.3cm}p{0.3cm}p{0.3cm}}	\hline
			\multicolumn{1}{c}{\multirow{3}{*}{Attack}} 	& \multicolumn{8}{c}{Reference and Year} \\ \cline{2-9} 
															& \cite{36} & \cite{13} & \cite{12} & \cite{8}  & \cite{34} & \cite{44} & \cite{49} & \cite{93}\\
															& 2015 		& 2016 		& 2016 		& 2017 		& 2018 		& 2019 		& 2019 		& 2020 \\ \hline
			\rowcolor[HTML]{EFEFEF} Re-compression          & Yes  		& Yes  		& Yes  		& Yes  		& Yes  		& Yes  		& Yes		& Yes\\
			\rowcolor[HTML]{FFFFFF} Cropping                & NM   		& NM   		& NM   		& NM   		& NM   		& NM   		& Yes 		& NM\\
			\rowcolor[HTML]{EFEFEF} Rotation                & NM   		& NM   		& NM   		& NM   		& NM   		& NM   		& Yes 		& NM\\
			\rowcolor[HTML]{FFFFFF} Scaling		            & NM   		& NM  		& NM   		& NM   		& NM   		& NM   		& NM		& Yes\\
			\rowcolor[HTML]{EFEFEF} Gaussian Noise          & NM   		& NM   		& Yes  		& Yes  		& Yes  		& Yes  		& NM		& Yes\\
			\rowcolor[HTML]{FFFFFF} Gaussian Filtering      & NM   		& NM   		& Yes  		& NM   		& Yes  		& Yes  		& NM		& NM\\
			\rowcolor[HTML]{EFEFEF} Salt and Pepper         & NM   		& NM   		& Yes  		& Yes  		& NM   		& Yes  		& NM		& Yes\\
			\rowcolor[HTML]{FFFFFF} Frame Delete/Replace 	& NM   		& Yes  		& NM   		& NM   		& Yes  		& NM  		& NM		& NM\\
			\rowcolor[HTML]{EFEFEF} AWGN Noise              & NM   		& Yes  		& NM   		& NM   		& NM   		& NM   		& NM		& NM\\
			\rowcolor[HTML]{FFFFFF} Rayleigh Fading Noise   & NM   		& Yes  		& NM   		& NM   		& NM   		& NM   		& NM		& NM\\ \hline
			\multicolumn{9}{c}{NM = Not Mentioned, AWGN = Additive White Gaussian Noise}
		\end{tabular}}
	\end{center}
\end{table}

In 2015, reference \cite{36} proposed a blind embedding-based watermarking technique, which alters the NNZ coefficients of 4x4 transform blocks of the HEVC video sequence. Simulation results showed the robustness of this technique against re-compression attacks. However, it suffered from bitrate increase issue due to altering zero coefficients to NZ, which negatively affected the BIR and the visual quality of the HEVC video.

In 2016, reference \cite{13} applied the repetition and the Bose-Chaudhuri-Hocquenghem (BCH) codes to detect and correct errors in order to improve the robustness of the proposed watermarking technique against the common attacks. However, this approach introduced high complexity that negatively affects the total performance of the watermarking process. 

Later, Dutta et al. proposed two embedding-based approaches \cite{12, 8} that embed the watermark in stable zones of the I-frames and P-frames, respectively. The main shortage of these two techniques was their sensitivity to synchronization and drift errors. For this reason, the authors decided to send the location map (\textit{palette}) to the decoder side which increased the robustness of these techniques at the expense of transmission overhead.

In 2018, reference \cite{34} proposed an efficient drift-free embedding-based watermarking technique using the stable zones of the HEVC I-frames. Experimental results reveal the robustness against re-compression, noise, and temporal attacks while maintaining the visual quality and the bitrate increase. However, its robustness is still low under the frame dropping, and Gaussian filter and noise attacks compared to the recent references which were not taken into this work account, where it was only compared to \cite{36,12} published in 2015 and 2016, respectively. Moreover, it also neglected the motion characteristics during the selection of embedding zones, which adds the possibility of being affected by synchronization errors.

In 2019, reference \cite{44} proposed a new technique that embeds the watermark into the intra-prediction residual 4x4 blocks at the spatial domain. This approach showed an acceptable level of robustness due to the avoidance of distortion drift errors. However, this technique is not robust enough against the frame dropping/exchange attack, for this reason, it is highly recommended to use the Error detection and correction codes to strengthen it against such attacks.

In the same year, reference \cite{49} proposed a robust video zero-watermarking technique based on the discrete wavelet transform and singular value decomposition to solve the problems associated with Reference \cite{101} such as false detection bits and visual distortion. This technique showed high robustness against re-compression, common image processing attacks, and geometric attacks. However, this technique is highly complex due to employing hybrid transforms based on the discrete wavelet and bi-orthogonal transform, and singular value decomposition. Moreover, it needs to improve its resistance against high-intensity rotation attacks, and also it has to eliminate the false-alarm problem. Additionally, this technique is not done during the encoding/decoding of the HEVC video, which requires extra processing before and after the encoding and decoding, respectively.

Recently in 2020, reference \cite{93} uses the stable regions to embed the invariant features, where simulation results showed high resistance to some attacks; such as noise, color, and brightness correction of frames, and scaling. However, the embedding capacity of the proposed algorithm was very limited, which significantly affects the recovery quality of the watermark and the PSNR value if the size of the embedded watermark increases. Moreover, it could be fragile against the frame dropping/exchange attack, for this reason, it is highly recommended to apply the error detection and correction codes to increase its robustness.

\subsubsection{\textbf{Security challenges:}}
Most existing watermarking techniques rely on some operations and rules that could make the extraction process an easy task if these operations and rules become known to the public. In order to solve this issue, some researchers went to use randomization functions in their embedding/extraction operations and rules while some used encryption algorithms to secure the watermark information.

References \cite{8,12} propose techniques for security based on random functions of embedding blocks selection without encrypting the embedded watermarks. Consequently, if those random functions became known to the public, the watermarks could be easily retrieved by attackers.

Since extracting watermarks is not needed for authentication applications, irreversible cryptography techniques, such as MD5, SHA-1, and SHA-256, are highly recommended to detect unauthorized tampering of HEVC videos. Accordingly, references \cite{1,14} have applied SHA256 hash functions on the extracted features from HEVC videos to prevent code imitation for authenticated video. Contrarily, reference \cite{93} has applied the Arnold scrambling transform for authentication purposes which is a reversible technique not strong enough to be used for such application.

For copyright applications, the receiver side has to retrieve the embedded watermark from HEVC video, that is why reversible cryptography techniques are highly recommended for such application \cite{41}. Reference \cite{13} implemented an extended Arnold scrambling transform, for copyright application, to encrypt the watermark before embedding it into HEVC video, then, to decrypt it after the extraction process at the receiver side.

Indeed, using cryptography techniques for HEVC codec may receive high interest from researchers in this area to improve the capability of current watermarking techniques at preventing attackers from accessing watermarks information. However, HEVC designers have to balance between security and complexity to increase the practicality of implementing video codecs, especially if real-time video streaming is targeted.

\subsection{Q8: Are there any real-time HEVC video watermarking techniques implemented on hardware platform?}
Even though most of the existing software-based watermarking solutions are able to perform the embedding and extraction processes on HEVC videos, they are still considered time-consuming solutions due to complexity that affect their applicability for real-time applications. 

In fact, these time-consuming solutions could give enough chance to the attackers to attack the HEVC video and to manipulate its content. Thus, it is crucial to develop hardware-based solutions for real-time HEVC watermarking, where watermarks data can be embedded and extracted in a timely manner.

Recently, references \cite{43,47} proposed two real-time HEVC watermarking techniques for ownership verification based on Very-Large-Scale Integration (VLSI) architecture and it was applied on a Field Programmable Gate Array (FPGA) platform. Both techniques use AC-NZ coefficients at the HEVC entropy part for embedding and extracting watermarks, while they control the bitrate increase and minimize the complexity. However, these two techniques suffer from slightly high drift and synchronization errors due to the CABAC open loop that lacks feedback to minimize errors. Moreover, using this option is fragile to some extent and sensitive to the common attacks such as re-compression and image processing attacks, which made it more applicable for authentication applications. For more details, refer to Q3-1.1 and Q4-Option III.

Finally, designing and implementing HEVC hardware-based watermarking techniques that are able to perform properly for real-time applications is still an open challenge. Moreover, hardware-based error detection and correction codes, such as BCH code, could be used with such systems to improve efficiency and robustness to give further support to copyright applications.

\section{Conclusion} \label{conc}
This SLR has been conducted to explore the current status of the research about HEVC watermarking techniques for authentication and copyright purposes. The main aim of this paper is to collect the related resources and references according to a clear process and specific rules in order to identify the challenges and open issues on the HEVC video watermarking. Additionally, It identifies out the potential research directions for interested researchers and developers. 

The time scope of this SLR covers all research articles published in the period of time from January 2014 up to the end of April 2020, where 343 articles published in this area have been found and only 42 articles have met the selection criteria. Then,  the selected 42 articles have been analyzed and discussed to identify the challenges of adopting HEVC  watermarking techniques to support authentication and copyright purposes. Moreover, a new classification for the existing up-to-date techniques has been drawn based on many factors, such as the domain, extraction and embedding process, human perceptibility, and fragility and robustness factors. Thereafter, this classification has been discussed to give a clear view to researchers who are interested in this area.

Eventually, we dedicate this important reference for researchers who are interested in this area, which will save their time and effort by facilitating the process of finding related works and references. Additionally, this paper summarizes the existing literature and highlights the main challenges and open issues.

\section{Future Directions} \label{futur}
In the future, interested researchers on the HEVC watermarking techniques for authentication and copyright application should consider the followings:
\begin{itemize}
	\item The watermark zone selection criteria should be carefully set to fit the needs of authentication and copyright applications, which by its role could minimize the bitrate increase and minimize the drift and synchronization errors.
	\item The error detection and correction codes, such as BCH, could be considered to reduce most types of errors when high-robustness watermarking is required.	
	\item All potential attacks that could affect the watermark readability have to be taken into account to allow the watermarks to stand strongly against them when high-robustness watermarking is required.
	\item There should be a great balance between complexity and security.
	\item Hardware-based solutions to support real-time applications should be considered.
	\item It might be a great idea if we take the Versatile Video Coding (VVC) features and HEVC features into account when developing watermarking techniques to make it ready for the VVC standard.
\end{itemize}




\begin{IEEEbiography}[{\includegraphics[width=1in,height=1.25in,clip,keepaspectratio]{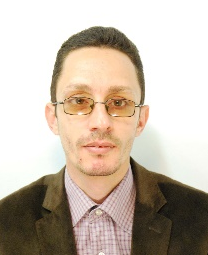}}]{Ali A. Elrowayati}
	(M'16) received his bachelor in Electronic Engineering from the College of Industrial Technology, Misurata, Libya in 2001, and his Master in Electrical Engineering from Faculty of Engineering, Misurata University, Libya in 2011. Currently, he is a Ph.D. candidate in the Faculty of Electrical and Electronic Engineering, Universiti Tun Hussein Onn Malaysia. Currently, he is working on a High-Efficiency Video Watermarking project for Copyright Protection and Authentication support. His research interests are image processing, video coding, watermarking, and neural network.
\end{IEEEbiography}
\vfill
\newpage

\begin{IEEEbiography}[{\includegraphics[width=1in,height=1.25in,clip,keepaspectratio]{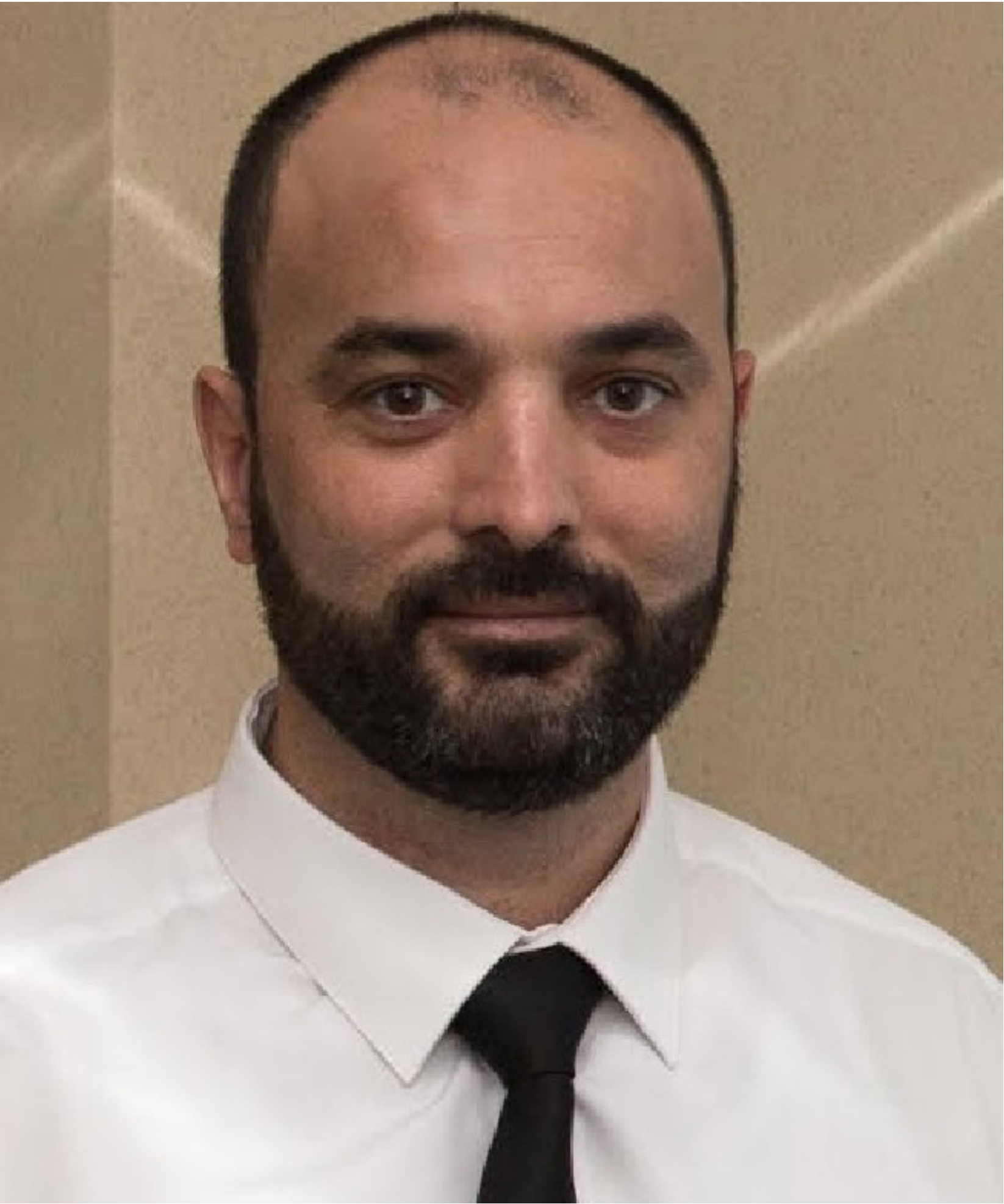}}]{Mohamed A. Alrshah}
	(M'13--SM'17) received his BSc degree in Computer Science from Naser University - Libya, in 2000, and his MSc and Ph.D. degrees in communication technology and networks from Universiti Putra Malaysia in May 2009 and Feb 2017, respectively. Now, he is a Senior Lecturer in the Department of Communication Technology and Networks, Faculty of Computer Science and Information Technology, Universiti Putra Malaysia (UPM). Also, he is a senior member of the IEEE. He has published a number of articles in high impact factor scientific journals. His research interests are in the field of high-speed TCP protocols, high-speed wired and wireless networks, parallel and distributed algorithms, WSN, IoT, and cloud computing.
\end{IEEEbiography}

\begin{IEEEbiography}[{\includegraphics[width=1in,height=1.25in,clip,keepaspectratio]{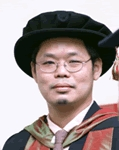}}]{Mohammad Faiz Liew Abdullah}
	(M'01--SM'14) received BSc (Hons) in Electrical Engineering (Communication) in 1997, Dip Education in 1999, and Master of Engineering in Optical Fiber Communication in 2000 from University of Technology Malaysia (UTM). He completed his Ph.D. in August 2007 from The University of Warwick, the United Kingdom in Wireless Optical Communication Engineering. He started his career as a lecturer at Polytechnic Seberang Prai (PSP) in 1999 then he has been transferred to UTHM in 2000. Currently, he is a Professor in the Department of Communication Engineering, Faculty of Electrical \& Electronic Engineering, University Tun Hussein Onn Malaysia (UTHM). He had 18 years experience of teaching in higher education, which involved the subject Optical Fiber Communication, Advanced Optical Communication, Advanced Digital Signal Processing and etc. His research interests are Wireless and Optical Communication, Solar Cell Fabrication, Image Watermarking, and Robotics in communication.
\end{IEEEbiography}

\begin{IEEEbiography}[{\includegraphics[width=1in,height=1.25in,clip,keepaspectratio]{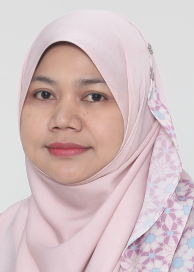}}]{Rohaya Latip}
	(M'01) is an Associate Professor at the Faculty of Computer Science and Information Technology, University Putra Malaysia. She holds a Ph.D. in Distributed Database and Msc. in Distributed System from University Putra Malaysia. She obtained her Bachelor of Computer Science from University Technology Malaysia, Malaysia in 1999. She is currently Head of the Department of Communication Technology and Network. Her research interests include Big Data, Cloud and Grid Computing, Network management, and Distributed database. She was also Head of the HPC section in University Putra Malaysia (2011-2012) and consulted the campus grid project and also the wireless project for the hostel of the UPM campus. She is also a Co-researcher at Institute for Mathematics Research (INSPEM). 
\end{IEEEbiography}
\EOD
\vfill
\end{document}